\documentclass[12pt,a4paper]{article}
\usepackage{color}
\usepackage{cite}
\usepackage{epsfig}
\usepackage{amssymb}
\usepackage{amsmath}

\usepackage{bm}                 
\usepackage{cancel}
\usepackage{comment} 
\newcommand{\pres}[2]{\setbox0=\hbox{$\scriptstyle #1$} \dimen0=\dp0  
              \dimen1=\ht0 \divide\dimen1 by 3
              \advance\dimen0 by \dimen1
              \hbox{\lower\dimen0 \hbox{$\scriptstyle #2\!\!$}} #1}
\newcommand{\nc}{\newcommand}
\nc{\figcap}[1]{\begin{quote}\refstepcounter{figure}
        {\bf Figure \thefigure}: {\small #1}\end{quote}}
\nc{\fig}[1]{\mbox{Fig.~\ref{#1}}}
\nc{\noi}{\noindent}
\nc{\bea}{\begin{eqnarray}}
\nc{\eea}{\end{eqnarray}}
\nc{\bean}{\begin{eqnarray*}}
\nc{\eean}{\end{eqnarray*}}
\nc{\ba}{\begin{array}}
\nc{\ea}{\end{array}}
\nc{\be}{\begin{equation}}
\nc{\ee}{\end{equation}}
\nc{\nn}{\nonumber}
\nc{\bra}[1]{\langle #1|}
\nc{\ket}[1]{|#1\rangle}
\nc{\av}[1] {\langle #1\rangle}
\nc{\vac}[1] {\langle 0| #1|0\rangle}
\nc{\amp}[2]{\langle #1|#2\rangle}
\nc{\da}{\dagger}
\nc{\pa}{\partial}
\nc{\ga}{\gamma}
\nc{\ep}{\epsilon}
\nc{\tf}{t_f}
\nc{\half}{\ensuremath{\frac{1}{2}}}
\nc{\hHH}{\hat H}
\nc{\ha}{\hat a}
\nc{\hO}{\hat O}
\nc{\hAA}{\hat A}
\nc{\hB}{\hat B}
\nc{\hG}{\hat G}
\nc{\hN}{\hat N}
\nc{\hU}{\hat U}
\nc{\hx}{\hat{x}}
\nc{\hp}{\hat{p}}
\nc{\hpsi}{\hat \psi}
\nc{\hphi}{\hat \phi}
\nc{\hpi}{\hat \pi}
\nc{\hpd}{\hat \psi ^\dagger}
\nc{\hE}{\hat E}
\nc{\hb}{\hat b}
\nc{\hc}{\hat c}
\nc{\hjo}{\hat j _0}
\nc{\hrho}{\hat \rho}
\nc{\leave}{\! \! \! \! \! / \, \,}
\nc{\intl}[1]{\int d\! #1 \,} %defines a good spacing for integrals
\nc{\intll}[3]{\int _#1^#2 d\! #3 \,} % integral with limits
\nc{\lm}{\lim _{y \rightarrow x}}
\nc{\scd}{\partial ^2 _{A_T}}
\nc{\fd}[1]{\frac{\delta }{\delta #1}} % functional derivative
\nc{\pad}[1]{\frac{\partial}{\partial #1}} % partial derivative
\nc{\refpa}[1]{(\ref{#1})} % referencing equations with brackets
\nc{\calH}{\ensuremath{\mathcal{H}}}
\nc{\calD}{\ensuremath{\mathcal{D}}}
\nc{\calL}{\ensuremath{\mathcal{L}}}
\nc{\calO}{\ensuremath{\mathcal{O}}}
\nc{\hcalO}{\ensuremath{\hat \mathcal{O}}}
\nc{\calK}{\ensuremath{\mathcal{K}}}
\nc{\Tr}{\ensuremath{\mathrm{Tr}}}
\nc{\tr}{\ensuremath{\mathrm{tr}}}
\nc{\ra}{\rightarrow}
\nc{\lr}{\leftrightarrow}
\nc{\phistar}{\phi^*}
\nc{\etat}{\eta_T}
\nc{\het}{\hat E_T}
\nc{\hpt}{\hat \psi_T}
\nc{\hpdt}{\hat \psi ^\dagger_T}
\nc{\bart}{\bar{t}}
\nc{\barp}{\bar{p}}
\nc{\barT}{\bar{T}}
\nc{\hbarrho}{\hat{\bar{\rho}}}
\nc{\bga}{\ensuremath{\mbox{\boldmath{$\gamma$}}}}
\nc{\bsi}{\ensuremath{\mathbf{\sigma}}}
\nc{\bx}{\ensuremath{\mathbf{x}}}
\nc{\by}{\ensuremath{\mathbf{y}}}
\nc{\bz}{\ensuremath{\mathbf{z}}}
\nc{\bp}{\ensuremath{\mathbf{p}}}
\nc{\bn}{\ensuremath{\mathbf{n}}}
\nc{\bbp}{\ensuremath{\bar{\mathbf{p}}}}
\nc{\bP}{\ensuremath{\mathbf{P}}}
\nc{\hbA}{\hat{\ensuremath{\mathbf{A}}}}
\nc{\hbB}{\hat{\ensuremath{\mathbf{B}}}}
\nc{\bA}{\ensuremath{\mathbf{A}}}
\nc{\bJ}{\ensuremath{\mathbf{J}}}
\nc{\bB}{\ensuremath{\mathbf{B}}}
\nc{\bH}{\ensuremath{\mathbf{H}}}
\nc{\bM}{\ensuremath{\mathbf{M}}}
\nc{\bD}{\ensuremath{\mathbf{D}}}
\nc{\bE}{\ensuremath{\mathbf{E}}}
\nc{\hbE}{\hat{\ensuremath{\mathbf{E}}}}
\nc{\br}{\ensuremath{\mathbf{r}}}
\nc{\bj}{\ensuremath{\mathbf{j}}}
\nc{\bOm}{\ensuremath{\mathbf{\Om}}}
\nc{\om}{\omega}
\nc{\Om}{\Omega}
\nc{\sgn}{\mbox{sgn}}
\nc{\deltabar}{\mbox{$\delta\hspace*{-8pt}\vspace*{-8pt}-$}}
\nc{\gammat}{\tilde{\gamma}}
\nc{\mub}{\bar{\mu}}
\nc{\rhob}{\bar{\rho}}
\nc{\Bb}{\bar{B}}
\nc{\Jb}{\bar{J}}
\nc{\Mb}{\bar{M}}
\nc{\Tb}{\bar{T}}
\nc{\sbar}{\bar{s}}
\nc{\betab}{\bar{\beta}}
\nc{\hj}{\hat j}
\nc{\hQ}{\hat Q}
\nc{\hJ}{\hat J}
\nc{\hA}{\hat A}
\nc{\hH}{\hat H}
\nc{\de}{\delta}
\nc{\leri}{\leftrightarrow}
\nc{\llabel}[1]{\label{#1}\marginpar{#1}}
\nc{\Refer}[1]{Ref.\mbox{\hspace{0.5mm}\cite{#1}}}
\nc{\Referes}[1]{Refs.\mbox{\hspace{0.5mm}\cite{#1}}}
\nc{\bc}{\begin{center}}
\nc{\ec}{\end{center}}
\nc{\inv}[1]{\frac{1}{#1}}

\newlength{\overeqskip}
\newlength{\undereqskip}
\nc{\eq}[1]{\mbox{Eq.~(\ref{#1})}}
\nc{\eps}{\epsilon}
\nc{\goto}{\rightarrow}
\nc{\cF}{{\cal F}}
\nc{\cG}{{\cal G}}
\nc{\cH}{{\cal H}}
\newcommand{\Dlarge}[1]{\displaystyle #1}
\newcounter{sectionc}
\newcounter{subsectionc}
\newcounter{subsubsectionc}
\renewcommand{\thesubsectionc}{\thesectionc.\arabic{subsectionc}}  
\renewcommand{\thesubsubsectionc}{\thesubsectionc.\arabic{subsubsectionc}}     
\renewcommand{\section}[1]
{\refstepcounter{sectionc}\vspace{0.0cm}
\setcounter{subsectionc}{0}\setcounter{subsubsectionc}{0}\noindent 
	{\bf\thesectionc. #1}}
\renewcommand{\subsection}[1] {\vspace{0.0cm}
\refstepcounter{subsectionc} % Added by2026-01-08 by PK
\setcounter{subsubsectionc}{0}\noindent 
{\rm\thesubsectionc. #1}\par\vspace{0.4cm}}
      \renewcommand{\subsubsection}[1] {\vspace{0.6cm}\addtocounter{subsubsectionc}{1}
{\rm\thesectionc.\thesubsectionc.\thesubsubsectionc  #1}\par\vspace{0.0cm}}

\renewenvironment{thebibliography}[1]
	{\begin{list}{[\arabic{enumi}]}
	{\usecounter{enumi}\setlength{\parsep}{0pt}
\setlength{\leftmargin 0.52cm}{\rightmargin 0pt}
	 \setlength{\itemsep}{6pt} \settowidth
	{\labelwidth}{#1.}\sloppy}}{\end{list}}
\makeatletter
\newcommand{\setlabel}[1]{\edef\@currentlabel{#1}\label}
\makeatother
\begin{document}
% 
%\documeinstantaneouslyntclass[12pt]{article}
% 
% 
\setlength{\jot}{10pt} 
%
% ---------------------  headline ( start )  ------------------
% 
\thispagestyle{empty} 
\today  
\vspace*{10mm}
\begin{center}  
\baselineskip 1.5cm 
{
\Large\bf
 Fractional Angular Momentum and  Quasi-Probability Densities for Angular Degrees of Freedom
}
\\[5mm]  
\normalsize 
\end{center} 
\begin{center} 
\vspace*{5mm}
{\centering 
{\large Bo-Sture K. Skagerstam\footnote{Corresponding author}$^{,}$\footnote{Email address: bo-sture.skagerstam@ntnu.no}
\vspace*{1mm}
\\
Department of Physics \\  Norwegian University of Science and Technology - NTNU \\ N-7491 Trondheim,  Norway}}
\\
{\large and}
\\
{\centering 
{\large Per K. Rekdal\footnote{Email address: per.k.rekdal@himolde.no}
\vspace*{1mm}
\\
Molde University College\\  PO Box 2110, N-6402 Molde,  Norway}}
%
%}
%
%
%[1mm]
%
%
%\centering 
\end{center} 
% 
%
% ---------------------  headline  ( end )  ------------------
% 
% 
%
%\vspace{1cm}
%
%
%
% -----------------------   abstract  (start)  --------------------
%
\vspace*{1.0mm}
\begin{abstract} 
%
% \normalsize 
% 
\noindent In the present work we consider state-dependent quasi-probability  densities $W[\theta,p]$ and $W_{1/2}[\theta,p]$, depending on two  parameters $\theta$ and $p$, where  $-\pi \le  \theta \le \pi $  and   $p$   can take  any real number. For pure states and integer values of  $p$  the corresponding  marginal distributions for $W[p]$ and $W_{1/2}[p]$ are  positive and in accordance with Born's  rule in quantum mechanics for the angular momentum observable $L= -id/d\theta$. 
For the  states $\psi(\theta)$  of the form in \Refer{Skagerstam_2023}, that can have  half-integer angular momentum average  values,  negative  values of  $W[\theta,p]$ and $W_{1/2}[\theta,p]$ can  reveal   non-classical features  but in an   ambiguous manner.  In line with  \Refer{Padgett_2004}, it is shown that experimental data of the uncertainties $\Delta \theta$ and $\Delta L$ could be   sufficient  to reveal quantum-mechanical features  of such states $\psi(\theta)$  without necessarily making  use of  quasi-probability  densities.
\vspace{1mm}
\end{abstract} 
%
%
% -----------------------   abstract  (end)  --------------------
%
%%\begin{center}
%%%\today
%%
%%\end{center}
%
\vspace{0.5cm}
\newpage
\setcounter{page}{1}
%
%\seqnoll
%
%
\begin{center}
\section
{INTRODUCTION \label{sec:introduction}}
\end{center}

The famous Wigner distribution  \hspace{-1mm}\cite{Wigner_1932}, and related quasi-probability densities, play an important role in since they, e.g., can be used as  witnesses of genuine quantum-mechanical behaviour  and  in  the reconstruction of various quantum states from   experimental knowledge of them (see, e.g.,  \Referes{Connell_1983,Wigner_1984,Vogel_1989,Buzek_1995,Lee_1995,Barnett_1995,Kowalski_1996
,Leonhardt_1997,Schleich_2001,Davidovich_2002,Kenfack_2004,Paris_2004,Haroche_2006,Gross_2006,Soto_2008,Rigas_2011,
Kastrup_2016,Tilma_2016,Weinbub_2018,Haroche_2020,Vogel_2020,Kowalski_2021} 
for a very limited  set of contributions and reviews to this  field of research and references cited therein). 
 If such quasi-probability densities  show signs of negativity they cannot be interpreted as a classical probability distributions (see, e.g., \Referes{Hudson_1974,Barnett_1995,Leonhardt_1997,Schleich_2001,Kenfack_2004,Gross_2006,Vogel_2020})  as was also discussed in a somewhat different context on quantum computers by Feynman \hspace{-1mm}\cite{Feynman_1982}. 

%
%
%
%
%\noindent 

The original Wigner distribution \hspace{-1mm}\cite{Wigner_1932} is  unique (see, e.g., the informative discussion of  Ref. \cite{Davidovich_2002})  depending on two independent real parameters $p$ and $q$   related  to  a pair of canonical observable $Q$ and $P$ such that $[Q,P]=i$ in units of $\hbar =1$  and a conventional pure state $\ket{\psi}$,  and can be  defined in two equivalent manners, i.e., 
\begin{gather}  \label{eq:wigner2024_1}
    W[q,p] \equiv \frac{1}{\pi } \int_{-\infty}^{\infty} dl e^{\Dlarge{-i2pl}} \psi(q+l) \psi^*(q-l) = \frac{1}{2\pi } \int_{-\infty}^{\infty} dl e^{\Dlarge{-ipl}} \psi(q+l/2) \psi^*(q-l/2) ~ .
\end{gather}
Here $\psi(q) \equiv \langle q|\psi \rangle$, $\psi^*(q) \equiv \langle \psi|q \rangle$, and $Q\ket{q}=q\ket{q}$.  The definition of  $W[q,p]$ is, e.g., appropriate for  the quantum dynamics on  a real line or related single-mode optical quadratures in the field of quantum optics.  $W[q,p]$ has the positive marginal distributions $W[q]$  and $W[p]$ of the observables $Q$ and $P$, i.e., 
\begin{equation}  \label{eq:wigner2024_2}
 W[q]=\int_{-\infty}^{\infty}dp  W[q,p] = | \psi(q) |^2 \, \, , \,\,W[p]=\int_{-\infty}^{\infty}dq  W[q,p]= |\frac{1}{\sqrt{2\pi} }\int_{-\infty}^{\infty}dqe^{\Dlarge{-iqp}} \psi(q)|^{\Dlarge{2}} \, ,
\end{equation}
which by construction are in accordance with Born's fundamental rule in quantum mechanics. The last expression in 
Eq.(\ref{eq:wigner2024_2}) involves the Fourier transform of the function $\psi(q)$ which in  Dirac notation corresponds to  $\psi(p) \equiv \langle p|\psi \rangle$  with  $P\ket{p}=p\ket{p}$ and using $\langle p|q \rangle = \exp (-ipq)/\sqrt{2\pi}$. We will encounter several  similar  notational  issues in  the present work. Various symbols must therefore, of course,  be understood  in the context in which they  occur. 

  The  experimentally well established     Yuen-Shapiro \hspace{-0.5mm}\cite{Yuen_1980} balanced  homodyne detection technique  in quantum optics  can, e.g., be used in order to  measure the Wigner function $W[q,p]$ above  (for a comprehensive  reviews on this and related  methods see, e.g., \Referes{Welsch_2009,Mauro_2003}). It is then  a  fundamental requirement  that  marginal distributions  as obtained in Eq.(\ref{eq:wigner2024_2})  
  are properly defined  probabilities in order to be  experimentally accessible.

\indent In physical situations with discrete quantum numbers one may consider embeddings into continuous degrees of freedom as  in  Sec.IV in \Refer{Preskill_2001}.  Here we will  focus our attention on  a specific pure quantum state $\psi(\theta)$,  previously investigated in  \Refer{Skagerstam_2023}, and 
investigate  to what extent   alternative forms of quasi-probability densities  can be used  to reveal  quantum mechanical  features of $\psi(\theta)$  and, in particular, the appearance of  fractional expectation  values of the angular momentum $L$.
    The construction of the corresponding  quantum state  $\psi(\theta)$ in \Refer{Skagerstam_2023} was motivated  by  various experimental procedures to generate restricted values of  angular position and angular momentum degrees of freedom and, in particular, the work in \Refer{Padgett_2004} (also see \Referes{Padgett_2005,Barnett_2007, Padgett_2010} and references cited therein). 

The  presence  of integer and  fractional angular momentum  vortex structures of optical light beams  has been  investigated  in great detail in the literature (see, e.g., \Referes{Leach_2002,Leach_2004,Berry_2004,Woerdman_2005,Tao_2005,Yao_2006, Gotte_2007,Tanimura_2015,Wang_2015,Balantine_2016,Mitri_2016,Pan_2016,
Deng_2019,Huang_2019,Chen_2021, Guo_2021, Deach_2022,Wang_2022,Liang_2023}). The degrees of freedom  then correspond to a transverse and  periodic angle $\theta$ and a corresponding angular momentum observable $L$.  

There are, of course, infinitely many choices of quasi-probability distributions  at hand as, e.g., considered in \Referes{Widder_54,Cahill_1969,KlauderSkagerstam_2007}.  
In the present paper, with a  special attention on  quantum states with fractional  expectation values of angular momentum $L$,  
 we will restrict ourselves to two  different, but  nevertheless quite  natural extensions $W[\theta,p]$ and $W_{1/2}[\theta,p]$  of the  original Wigner distribution   as defined by Eq.(\ref{eq:wigner2024_1})  to the case of  a finite domain of $\theta$.  They   lead, however,   to ambiguous but related physical conclusions. In terms of physical degrees of freedom these quasi-probability  densities are, e.g.,  such that their  domains of visibility and negativity do not necessarily overlap and that both of them can, in fact, be in conflict with Born's  rule in quantum mechanics if the parameter $p$ is not integer valued.
 
 We will not discuss to what extent the quasi-probability  densities $W[\theta,p]$ and $W_{1/2}[\theta,p]$ can  be  determined experimentally but limit ourselves to exhibit features of them indicating quantum-mechanical properties of the particular  state  $\psi(\theta)$ under consideration.
 It will, however, also be argued that for a fractional expectation value of the observable $L$,  special features of  the quantum state $\psi(\theta)$ that we consider   could be revealed  directly from a  knowledge of measurable uncertainties $\Delta \theta$ and $\Delta L$ without resorting to the general properties of quasi-probability densities following the experimental procedure  in \Refer{Padgett_2004}.

The paper is organized as follows. In Section  \ref{section_our_state} we recall various expressions for the parametric representation of the $2\pi$-periodic  state $\psi(\theta)$  of  \Refer{Skagerstam_2023} and present explicit expressions  for expectation values $\langle L \rangle$  and the  uncertainties $\Delta L$ with a special focus  on factional values of $\langle L \rangle$.  In Section  \ref{section_ourW} we present one  form of a quasi-probability distribution $W[\theta,p]$  for a finite domain $-\pi \le  \theta \le \pi$ with marginal distributions  for  the parameters $\theta$ and $p$. The  marginal distribution  $W[p]$  as obtained from $W[\theta,p]$  and for   integers $p$ is shown to be in accordance with   Born's  rule in quantum mechanics. This quasi-probability function $W[\theta,p]$ has a natural extension to any real number $p$. The  corresponding distribution  $W[\theta]$  is  carefully evaluated making use of the Fej\'{e}r summation method for integer-valued $p$, or  Ces\`{a}ro summability for integrals  when $p$   can be  any real number,  taking  boundary conditions at $\theta = \pm \pi$ into account.  
In Section \ref{section_polish_W} we recall   another definition of a quasi-probability density $W_{1/2}[\theta,p]$  as discussed in \cite{Kastrup_2016,Kowalski_2021} with a marginal  distribution $W_{1/2}[p]$  which  also is consistent with  Born's  rule  for integers $p$.  For arbitrary real values of $p$  the quasi-probability distributions $W[\theta,p]$  and  $W_{1/2}[\theta,p]$  are in general different and lead to different marginal distributions.  The methods described  in  Section \ref{section_ourW} are used to find,  in a precise manner, the corresponding marginal distributions $W_{1/2}[p]$  and $W_{1/2}[\theta]$  which turn out to be different  from  $W[p]$ and $W[\theta]$.
In Section \ref{section_our_uncertainties} we extend the results of  Ref.\cite{Padgett_2004}   for  the uncertainties $\Delta \theta$ and $\Delta L$  using the $2\pi$-periodic state $\psi(\theta)$ of \Refer{Skagerstam_2023} with a special focus on fractional values of the parameter $p$.  Finally, in Section \ref{section_our_conclusions} we summarize our findings.

The presentation of this work is aimed  to be self-contained. Various mathematical methods made use of are  outlined and the  readers are provided with a set  of  selected references for  more detailed expositions of the corresponding mathematical techniques.

\begin{center}
\section{CHOICE OF QUANTUM STATE \label{section_our_state}}
\end{center}

\begin{center}
\subsection{Extension of the S. Franke-Aronold et al.   2004  
  Intelligent  State  \cite{Padgett_2004}\label{sec:subsection_extension_of_state}
}  
\end{center}

 \indent The  pure quantum state $\psi(\theta)$ which  we, in particular, will consider   taking  limitations of degrees of freedom  into account can be expressed in terms of a  well-defined  $2\pi$-periodic  extension of the  state  defined on a circle  with an angular position $\theta$ in the range $-\pi \le \theta \le \pi$ as discussed in detail in \Refer{Padgett_2004}. In terms of a parameter $\lambda > 0$,   its  convergent Fourier series expansion reads 
\begin{gather} 
\psi(\theta)   = \sum_{n=-\infty}^{\infty}c(n)\psi_n(\theta)=  \frac{N}{\sqrt{\lambda}}\sum_{n=-\infty}^{\infty}e^{\displaystyle{-in\bar{\theta}}}e^{ -\displaystyle{(n - \bar{l})^2/2\lambda}} \psi_n(\theta)\, \,  \nonumber \\
= \frac{N}{\sqrt{2\pi\lambda}} e^{ \displaystyle{il(\theta}- \bar{\theta})}\sum_{n=-\infty}^{\infty} e^{ \displaystyle{in(\theta}- \bar{\theta})}e^{ -\displaystyle{(n - \eps)^2/2\lambda}} \, \, , 
 \label{eq:wigner2024_3}
\end{gather}
where $\psi_n(\theta)= \exp(in\theta)/\sqrt{2\pi}$ is a  normalized orthonormal  eigenfunction of the angular momentum operator $L= -id/d\theta$.  The Fourier coefficients $c(n)$ have here been identified in accordance with   the  state in  \Refer{Skagerstam_2023}. The space of parameters is therefore given by the set $\Omega = \{ \bar{\theta},\lambda ,\bar{l}\}$. We will  treat $\bar{\theta}$ as a free parameter often  with the choice  $\bar{\theta}=0$ for  reasons of simplicity unless explicitly specified. For arbitrary values of the real-valued parameter  $\bar{l}$  the  last expression in Eq.(\ref{eq:wigner2024_3})  is expressed in terms of $\bar{l} = l + \eps$, where $l$ is an integer and  $0 \leq \eps <1$. 
The last  expression in Eq.(\ref{eq:wigner2024_3}) will  be used in sections below   for small  $\lambda$ considerations.  

In passing, we observe that a completely different,  but formally  related state  with a Gaussian-like  amplitude as in Eq.(\ref{eq:wigner2024_3}) as well as other related  states   have  been reviewed theoretically  as well as investigated experimentally  in \Refer{Soto_2008} (and references cited therein)   without considering the use of quasi-probability densities.

 The normalization factor $N$  in Eq.(\ref{eq:wigner2024_3}) is determined by the condition
\begin{gather} 
1= \int_{-\pi}^{\pi}d\theta |\psi(\theta) |^2 =
 \frac{N^2}{\lambda}\sum_{n=-\infty}^{\infty}e^{ -\displaystyle{(n - \eps)^2/\lambda}} 
 = \frac{N^2}{\lambda} e^{ \displaystyle{-\eps^2}/\lambda}\vartheta_3[ -i\eps/\lambda, e^{ \displaystyle{-1/\lambda}}]   \nonumber \\ 
=  N^2\sqrt{\frac{\pi}{\lambda}}\vartheta_3[ -\eps\pi, e^{ \displaystyle{-\lambda \pi^2}}]  \,\, ,
 \label{eq:wigner2024_4}
\end{gather}
independent of $\bar{\theta}$. Here we make use of properties of $\vartheta_3[z,q]$-functions under modular transformations
using the definition 
\begin{equation} 
	\vartheta_3[z,q] = \sum_{n=-\infty}^\infty q^{\displaystyle{n^2}} e^{\displaystyle{2niz}} ~ ,
 \label{eq:two_phase_6} 
\end{equation}
within the unit disk $|q|< 1$.
%
%
% ------------   figure  (start)   ---------------
\begin{figure}[htb]  
\vspace{-1.0cm}
\centerline{\includegraphics[width=16cm,angle=0]{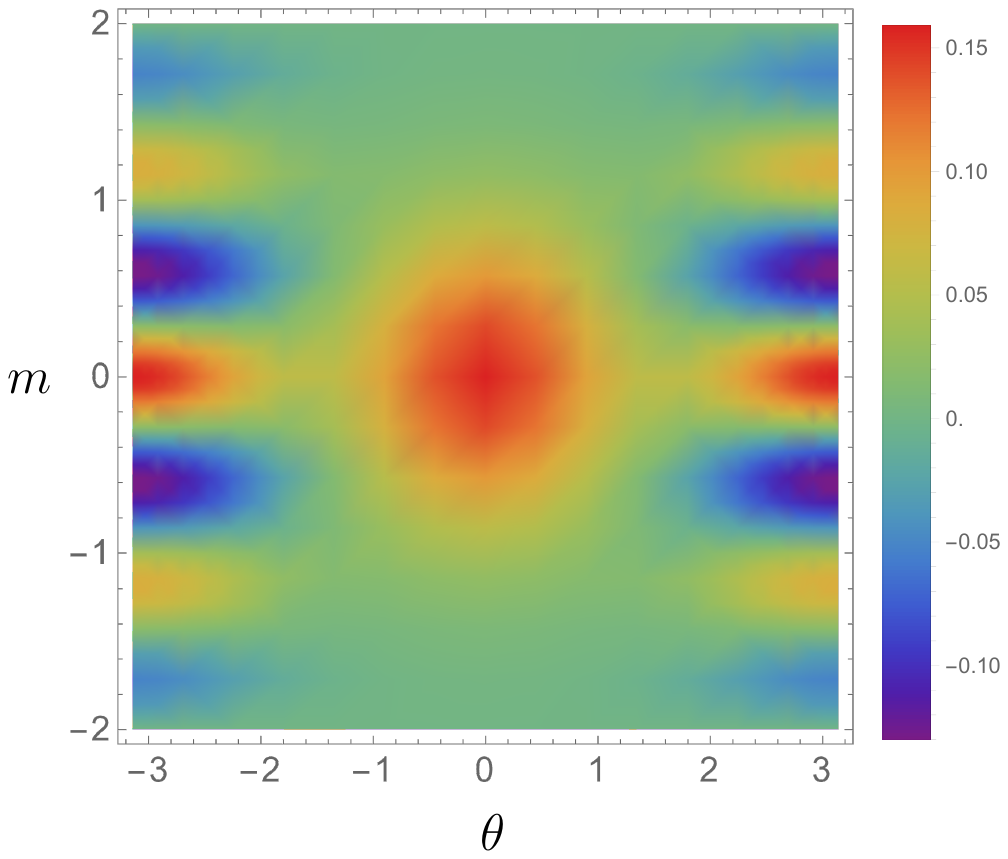}}
\vspace{-8.5cm}                 
\caption{The  distribution  $W[\theta,p]$ according to Eq.(\ref{eq:ourwigner_1})  with $-\pi \le \theta \le \pi$  and $\bar{\theta}=0$  for  the state $\psi(\theta)$  as given in  Eq.(\ref{eq:wigner2024_3}).
%In general  $W[\theta,p]$   is bounded  by  $|W[\theta,p]|\leq 1/2\pi$. 
Here   $q=\exp(-1/2\lambda)=0.5$ and $\bar{l}= l =\langle L \rangle$  integer-valued with  $W[\theta,m]$ as a function of the parameter  $m =p-l$ in  the range $-2\le m \le 2$. 
%instead as a function of $p$.  With this choice of parameters  
$W[\theta,m]$ is  then symmetric with respect to a line with $m=0$. For  half-integer values of $m$, in which case    the parameter  $p$ is  fractional,
%the Wigner distribution  
$W[\theta,m]$  can be negative at $|\theta| = {\cal O}(\pi)$.  
}
\label{fig:wignerplot_1}
\end{figure}
%
% ------------   figure   (end)    ---------------
%
%
 The normalization constant $N$ is therefore in general only a function of the parameters $\eps$   
 and the positive parameter  $\lambda$. For a given value of $\bar{l}$   the uncertainty  $\Delta \theta$ of  the quantum mechanical observable  $\theta$ is then an implicit function of the parameter $\lambda$  (details can be found in \Refer{Skagerstam_2023} for a more general  approach). 
 
  The state $\psi(\theta)$ as defined  in Eq.(\ref{eq:wigner2024_3}) can    also be expressed in terms of  the  
$\vartheta_3[z,q]$-function, i.e.,  
\begin{gather} 
\psi(\theta) =
%\frac{N}{\sqrt{2\pi\lambda}} e^{ \displaystyle{-\overline{l}^2 /2\lambda}  } \vartheta_3[( \theta -\bar{\theta})/2 -i %\overline{l}/2\lambda, q] \nonumber \\ =
 \frac{N}{\sqrt{2\pi\lambda}} e^{\displaystyle{il(\theta - \bar{\theta})}}e^{\displaystyle{- \eps^2/2\lambda}}  \vartheta_3[( \theta -\bar{\theta})/2 -i \eps/2\lambda, q]
 \label{eq:wigner2024_5}~ ,
\end{gather}
where $q=\exp(-1/2\lambda)$  and suitable for small $\lambda$ considerations. An alternative representation for $\psi(\theta)$ and suitable for large $\lambda$ is
\begin{gather} 
\psi(\theta) =N e^{ \displaystyle{i\overline{l}(\theta -\bar{\theta})}} e^{-\displaystyle{\lambda(\theta -\bar{\theta})^2/2}} 
 \vartheta_3[\pi\left(\eps +i\lambda(\theta-\bar{\theta})\right),e^{\displaystyle{-2\lambda\pi^2}}]
 \label{eq:wigner2004_5_1} ~ ,
\end{gather}
which easily  can be verified by making use of the Poisson summation method (see, e.g., Chapter $9$ in \Refer{Rudin_70} and \Refer{Schleich_1993}).

 When computing expectation values of $\theta$,  we  find the following explicit expression for $|\psi(\theta)|^2$  convenient in both  analytical  as well as numerical considerations
\begin{gather} 
 |\psi(\theta)|^2 =  \frac{|\vartheta_3 [\theta/2-i\eps/2\lambda,q]|^2}{2\pi\vartheta_3[-i\eps/\lambda,q^2]}
 ~ ,
     \label{eq:alternative_form}
\end{gather}
appropriate for small values of the parameter $\lambda$ where  $q=\exp(-1/2\lambda)$.  For large values of $\lambda$ we also  make use of the alternative expression
\begin{gather} 
 |\psi(\theta)|^2 = e^{\displaystyle{-\lambda\theta^2}} \sqrt{\frac{\lambda}{\pi}}
 \frac{|\vartheta_3 [\pi(\eps +i\lambda\theta),q^2]|^2}{\vartheta_3[-\eps\pi,q]}
%]|^2}{\vartheta_3[-i\eps/\lambda,q
%]
%}
 ~ ,
     \label{eq:second_ alternative_form}
\end{gather}
where  now $q=\exp(-\lambda\pi^2)$.

It is of importance to observe that $|\psi(\theta)|^2$ in these equations, and therefore also $\Delta\theta$,  do not depend on the integer $l$ in  $\bar{l}= l+\eps$. This  feature was also observed for the observable $\Delta L$   according to  Eq.(\ref{eq:wigner2024_4_2_3}).  The large or small values of the quantum number $l$ can therefore not  be used to distinguish between classical or quantum features in the expression  of  the probability distribution  $|\psi(\theta)|^2$ as long as we only consider the observables $\Delta\theta$ and $\Delta L$.

\begin{center}
\subsection{Expectation  Values of the Angular Momentum Observable $L$
\label{sec:subsection_expectation_values}}
\end{center}
 
The parameter $\bar{l}$ can now be related to various quantum mechanical expectation values of the observable $L$ such as
\begin{gather}
\langle L \rangle = \frac{N^2}{\lambda} \sum_{n=-\infty}^{\infty}ne^{\displaystyle{- (n-\overline{l})^2/\lambda}}   \,\,\,\, ,\,\, \,\,
 \langle L^2 \rangle = \frac{N^2}{\lambda} \sum_{n=-\infty}^{\infty}n^2e^{\displaystyle{- (n-\overline{l})^2/\lambda}} \,\, ,
 \label{eq:wigner2024_4_2}
\end{gather}
independent of the parameter $\bar{\theta}$ where, of course, $\langle {\cal O} \rangle = \langle \psi| {\cal O}|\psi\rangle$ for an observable ${\cal O}$ in quantum mechanics.  In terms of $\bar{l} = l + \eps$ as defined above,  we can write  Eqs.(\ref{eq:wigner2024_4_2}) in the following useful forms
 \begin{gather}
\langle L \rangle = l + \frac{N^2}{\lambda} \sum_{n=-\infty}^{\infty}ne^{\displaystyle{- (n-\eps)^2/\lambda}}   \,\,\,\, ,\,\, \,\,
 %\langle L^2 \rangle = \frac{N^2}{\lambda} \sum_{n=-\infty}^{\infty}n^2e^{\displaystyle{- (n-\overline{l})^2/\lambda}} \,\, ,
%
 \label{eq:wigner2024_4_2_1}
\end{gather}
and
 \begin{gather}
\langle L^2 \rangle = l^2 +\frac{N^2}{\lambda} \sum_{n=-\infty}^{\infty}n(n+2l)e^{\displaystyle{- (n-\eps)^2/\lambda}} \,\, .
 \label{eq:wigner2024_4_2_2}
\end{gather}
It then follows that
\begin{gather}
(\Delta L)^2 = \langle L^2 \rangle - \langle L \rangle^2= \frac{N^2}{\lambda}\sum_{n=-\infty}^{\infty}n^2e^{\displaystyle{- (n-\eps)^2/\lambda}} \nonumber \\
- ( \frac{N^2}{\lambda}\sum_{n=-\infty}^{\infty}ne^{\displaystyle{- (n-\eps)^2/\lambda}})^2 \ge 0 \, ,
 \label{eq:wigner2024_4_2_3}
\end{gather}
where the lower bound  can be  obtained directly by making use of  the Cauchy-Schwarz inequality. 
Therefore the expectation value $\langle L \rangle$ depends on the integer $l$ but the uncertainty $\Delta L$ depends only on the fractional part  $\eps$ of   $\bar{l} = l + \eps$.  

With $\eps = 0$  the integer $l$  can  be identified   with the  expectation value  $\langle L \rangle$ of the angular momentum.  The general expression for $(\Delta L)^2$  in Eq.(\ref{eq:wigner2024_4_2_3}) can be expressed in terms of the theta function  $\vartheta_3[z,q]$ and derivatives with respect to the parameter $z$. The corresponding large $\lambda$ asymptotic  expansions of these functions  can  be obtained  in a straightforward manner using Poisson summation techniques considered in a similar manner but in a different context  in \Refer{Itzykson_1973},  with, e.g.,  the result
\begin{gather}
(\Delta L)^2 = \frac{\lambda}{2}  - 2\pi^2\lambda^2e^{\displaystyle -\pi^2\lambda}\cos(2\pi\eps) + {\cal O}(\lambda^2\cos(4\pi\eps) e^{\displaystyle -2\pi^2\lambda})   >0 \, ,
 \label{eq:wigner2024_4_2_3_1}
\end{gather}
leading to  upper and lower bounds $\lambda /2 \pm  2\pi^2\lambda^2\exp( -\pi^2\lambda)$.  
In neglecting  the exponentially small terms ${\cal O}(\lambda^2 e^{\displaystyle -2\pi^2\lambda})$ one finds that numerically  Eq.(\ref{eq:wigner2024_4_2_3_1}) is  valid at least for $\lambda \ge 1$ 
   due to various non-trivial cancellations of $\eps$ and $\lambda$ dependent  terms.  These upper and  lower limits obtained  for $(\Delta L)^2$  are not necessarily the best  bounds but this issue will not be discussed in more detail here.  
 In the limit  of large values of $\lambda$, 
  %$q=\exp(-1/2\lambda) \rightarrow 1$ 
  we therefore  have that $\langle L \rangle = \bar{l}$.  By making use of the large $\lambda$-limit of   Eq.(\ref{eq:second_ alternative_form}) it then also follows that  the variance of  $\theta$ behaves like  $(\Delta \theta)^2 =1/2\lambda$   such that $\Delta \theta\Delta L =1/2$  in agreement   with \Referes
{Padgett_2004,Skagerstam_2023}. We remark that for some exceptional values of $\eps$ the terms indicated by ${\cal O}(\lambda^2 e^{\displaystyle -2\pi^2\lambda})$ in Eq.(\ref{eq:wigner2024_4_2_3_1}) should be replaced by even smaller terms. 

\begin{center}
\subsection{The Limit of Small $\lambda$ and Fractional Angular Momentum
\label{sec:subsection_expectation_values_2}}
\end{center}
In the limit of small $\lambda$ it  follows from Eq.(\ref{eq:wigner2024_4_2_3}) that 
\begin{gather}
(\Delta L)^2 =\frac{e^{ \displaystyle{-(1-2\eps)/\lambda}  } }{(1+e^{ \displaystyle{-(1-2\eps) /\lambda}  })^{\displaystyle{2}} } \, ,
 \label{eq:wigner2024_4_2_4}
\end{gather}
  which has an exponential sensitivity of  the parameter $\eps$. In particular we find a rather unique feature of the state  $\psi(\theta)$ as given in Eq.(\ref{eq:wigner2024_3}) namely  that $\Delta L \rightarrow 1/2$ as $\lambda \rightarrow 0$ when $\eps=1/2$ in which case the expectation value of  the angular momentum $L$ is fractional, i.e., $\langle L \rangle = l +1/2$ due to Eq.(\ref{eq:wigner2024_4_2_1}). 
 In the same limit of sufficiently small $\lambda$  and $\bar{l} = l + 1/2$ with  $l$ an integer,   it was also shown in \Refer{Skagerstam_2023}  that the state $\psi(\theta)$ takes the following  exact form in terms  of a superposition of angular momentum eigenstates 
\begin{gather} 
\psi(\theta) =\frac{1}{\sqrt{2}}\Huge( \psi_l(\theta)+ \psi_{l+1}(\theta) \Huge) \, , 
 \label{eq:lambda_zero}
\end{gather}
with $\bar{\theta}=0$ such that $\langle \theta \rangle =0$  and  $\langle L \rangle =l+1/2$,  and with uncertainties    $\Delta\theta =\sqrt{\pi^2/3-2}$ and  $\Delta L = 1/2$ that are independent of the integer $l$. It was also shown  in \Refer{Skagerstam_2023} that for any value of $\bar{l}$ arbitrarily close to but not equal  to a half-integer rational number we always obtain a branching to  angular momentum $l$ or $l+1$ eigenstates  for sufficiently small  but non-zero values of the parameter $\lambda$. For these  asymptotic eigenstates we have that  $\Delta\theta =\pi/\sqrt{3}$ and, of course,   $\Delta L =0$. We will return to these observations in more detail in Section \ref{section_our_uncertainties}.

A complete  set of admissible orthonormal fractional angular momentum eigenstates $\psi_{n\gamma}(\theta)$ of the operator $L$  with eigenvalues $n + \gamma$, which will be referred to in sections below, can easily  be constructed in the form
\begin{gather} 
\psi_{n\gamma }(\theta) =\frac{1}{\sqrt{2\pi}}e^{\displaystyle{i(n+ \gamma)\theta}}\, , 
 \label{eq:klauder_fractional}
\end{gather}
where   $n$ is an integer and where we always can choose $\gamma$ in the range $0 \leq  \gamma < 1$  (see, e.g., Theorem $4$ in Section $11.1$ in  Ref. \cite{Hellwig_1964} and for  more recent discussions  Ref.\cite{Geloun_2012}). Since trivially $|\psi_{n\gamma }(\theta)|^2= 1/2\pi$  we have that $\langle \theta \rangle =0$  and  $\Delta\theta =\sqrt{\pi^2/3}$ in contrast with the properties of the state $\psi(\theta)$ in Eq.(\ref{eq:lambda_zero}) with the smaller uncertainty $\Delta\theta =\sqrt{\pi^2/3-2}$. The probability $p_{l\gamma}$ to find the system in the state Eq.(\ref{eq:klauder_fractional}) with angular momentum $l+\gamma$ provided it has been prepared  in the state Eq.(\ref{eq:lambda_zero}) is given by $p_{l\gamma}=|\langle \psi_{l\gamma} |\psi\rangle|^2$ and one finds that  bounds
$1/2\leq p_{l\gamma} \leq 8/\pi^2$. We find that the maximal value of $p_{l\gamma}= 8/\pi^2$ actually corresponds to  the $\gamma = 1/2$ half-integer angular momentum  eigenstate  and that the minimal value $p_{l\gamma}=1/2$ corresponds to the angular momentum $\psi_{l}(\theta)$ or $\psi_{l+1}(\theta)$ eigenstates.
 %
 %
 %\begin{comment}
% ------------   figure  (start)   ---------------
\begin{figure}[htb]  
\vspace{-1.0cm}
\centerline{\includegraphics[width=16cm,angle=0]{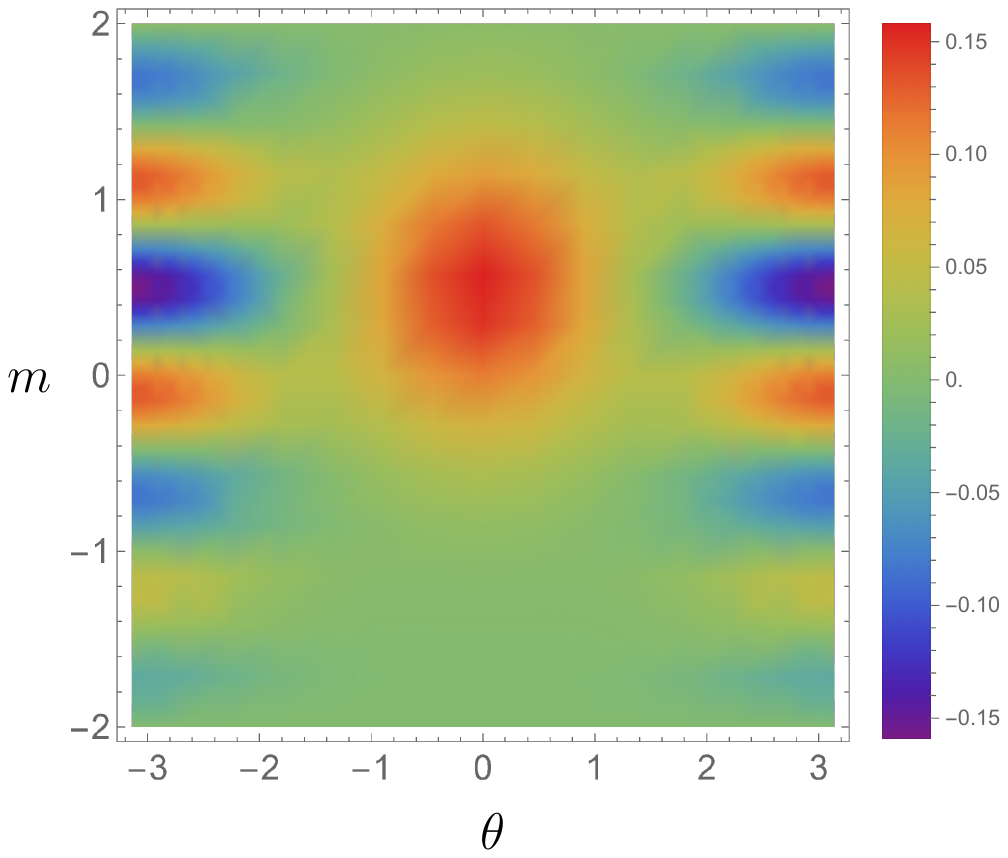}}
\vspace{-8.5cm}                 
\caption{The  distribution  $W[\theta,p]$ according to Eq.(\ref{eq:ourwigner_1})  with $-\pi \le \theta \le \pi$  and $\bar{\theta}=0$  for  the state $\psi(\theta)$  as given in  Eq.(\ref{eq:wigner2024_3}). 
%%In general  $W[\theta,p]$   is bounded  by  $|W[\theta,p]|\leq 1/2\pi$. 
Here  $q=\exp(-1/2\lambda)=0.5$ and a fractional  $\bar{l}= l+ 1/2$  with  $W[\theta,m]$ as a function of the parameter  $m =p-l$ in  the range $-2\le m \le 2$ instead as  a function of $p$.  The symmetry with respect to the line $m=0$ as in Fig.\ref{fig:wignerplot_1} is now not present.  For  half-integer values of $m$ the  distribution  $W[\theta,m]$  can be negative at $|\theta| = {\cal O}(\pi)$,  in which case    the parameter  $p$  is fractional. 
}
\label{fig:wignerplot_2}
\end{figure}
%
% ------------   figure   (end)    ---------------
%
%\end{comment}
 %
 %

%
%
%\newpage 
%
%
\begin{center}
\section{THE QUASI-PROBABILITY DISTRIBUTION  ${\bm{W[\theta,p]}}$
\label{section_ourW}}
\end{center}

\begin{center}
\subsection{ A Finite Domain for the Parameter  $\theta$
\label{subsection_Finite_Domain}}
\end{center}

 We are now interested in defining suitable quasi-probability  distributions in terms of two  parameters $\theta$ and $p$ such that $-\pi \leq \theta \leq \pi$ and where $p$ in principle can take any real value
 leading to  marginal distributions for the observables $\theta$ and $L$  from which expectation values can be computed  and compared with the rules of quantum mechanics for a  pure state $\psi(\theta)$ in a similar fashion as for the Wigner function $W[q,p]$  in Eqs.(\ref{eq:wigner2024_1}) and  (\ref{eq:wigner2024_2}).
The  quasi-probability  distributions we will consider  reveal the  presence of  fractional expectation values of  $\langle L \rangle$ for the  pure state $\psi(\theta)$ considered in \Refer{Skagerstam_2023}    for non-zero values of the parameter $\lambda$. 
 
 It is, however, now not obvious how to extend  the definitions in Eq.(\ref{eq:wigner2024_1})  appropriate  for quantum dynamics on a finite domain for the variable $\theta$  (for  reviews  and further references see, e.g., \Referes{Kastrup_2016,Tilma_2016,Kowalski_2021}). In the spirit of the first definition in Eq.(\ref{eq:wigner2024_1})
   we  propose the following real-valued definition of a quasi-probability  distribution  $W[\theta,p]$ which is $2\pi$-periodic in the variable $\theta$, i.e., 
\begin{gather}  \label{eq:ourwigner_1}
    W[\theta,p]= \int_{-\pi}^{\pi} \frac{d\theta'}{2\pi} e^{\Dlarge{-i2p\theta' }} \psi(\theta+\theta') \psi^*(\theta -\theta')  = \int_{-2\pi}^{2\pi}  \frac{d\theta'}{4\pi} e^{\Dlarge{-ip\theta' }} \psi(\theta+\theta'/2) \psi^*(\theta -\theta'/2)~ ,
\end{gather}
with an integrand which is also  $2\pi$-periodic in $\theta'$ for integer values of the parameter $p$. The range of integration in Eq.(\ref{eq:ourwigner_1})  is chosen in such a way that the  entire domain $\theta \in [-\pi,\pi]$ of periodicity for  the state $\psi(\theta)$ is taken into account.  As  will be argued below this range of integration is of particular importance when we consider a quantum state  with fractional angular momentum.  The second equivalent expression in Eq.(\ref{eq:ourwigner_1}) will be referred to below. This  distribution  $ W[\theta,p]$ is  bounded according to  $|W[\theta,p]|\leq 1/2\pi$   as follows from the Cauchy-Schwarz inequality.

In  terms of an integer $l$ and  $0\leq \eps <1$ and  $\bar{l} = l + \eps$, it can now be verified that
the   distribution $W[\theta,p]$ according to Eq.(\ref{eq:ourwigner_1})  can be expressed  in the following general  form
\begin{gather} 
    W[\theta,p] =  \frac{1}{2\pi\vartheta_3[-i\eps/\lambda,q^2]} \nonumber \\ \times \int_{-\pi}^{\pi} \frac{d\theta'}{2\pi}e^{\Dlarge{-i2m\theta' }}\vartheta_3[\frac{\theta + \theta'}{2}  - i\eps/2\lambda,q]\vartheta_3[\frac{\theta - \theta'}{2}  +i\eps/2\lambda,q] ~ ,
     \label{eq:integerlbar}
\end{gather}
with parameters $m=p-l$ and  $q=\exp(-1/2\lambda)$. We have found  that Eq.(\ref{eq:integerlbar}) turns out to a very  useful  form of $W[\theta,p]$ in many of our numerical considerations. Here we observe the simple but useful identity
\begin{gather} 
 \vartheta_3^*[\frac{\theta - \theta'}{2}  -i\eps/2\lambda,q]= \vartheta_3[\frac{\theta - \theta'}{2}  +i\eps/2\lambda,q] =\vartheta_3[\frac{\theta'  -\theta}{2}  -i\eps/2\lambda,q]~.
 \end{gather}

In Fig.\ref{fig:wignerplot_1} we show  the behaviour of the quasi-probability  distribution  $W[\theta,p]$  for the state  in Eq.(\ref{eq:wigner2024_3})   when  $q=\exp(-1/2\lambda)=0.5$ and $\bar{l}=l =\langle L \rangle$ is  an integer $l$ as a function of $m=p - l$ instead as a function of $p$. For   other values of $q$ such that $q  \geq 0.5$   the presence  of  negativity of  the corresponding distribution $W[\theta,m]$ remains essentially the same as illustrated in Fig.\ref{fig:wignerplot_1}. Even if $\langle L \rangle$ in this case is an integer we observe that  $W[\theta,m]$  is negative  for fractional values of $p$ in  the vicinity of  $p=l \pm 1/2$.

    In a similar manner as in Fig.\ref{fig:wignerplot_1} we show in Fig.\ref{fig:wignerplot_2}  with  $q=\exp(-1/2\lambda)=0.5$ the behaviour of the  distribution $W[\theta,p]$  for the state  in Eq.(\ref{eq:wigner2024_3})  using  $\bar{l}=l+1/2$ as a function of $m=p - l$ instead as a function of $p$. For   other values of $q$ such that $q \geq 0.5$   the presence  of  negativity of  the corresponding   distribution $W[\theta,m]$ also remains essentially the same as illustrated in Fig.\ref{fig:wignerplot_2}.   In the limit as $q \rightarrow 1$  we have the  expectation value  $\langle L \rangle = l+1/2$.  
    %This is an indication that the Wigner function $W[\theta,p]$  is sensitive to the presence of  a fractional average %values of $\langle L \rangle$.

In Fig.\ref{fig:wignerplot_2} the corresponding  distribution $W[\theta,p]$  in the form  of   Eq.(\ref{eq:ourwigner_1}) can  clearly be  negative for fractional  values  in the vicinity of $p=l \pm 1/2$  at $|\theta| = {\cal O}(\pi)$.  We observe, in particular,  that for the fractional value  $m=1/2$, i.e., $p=l + 1/2$,  the presence of negativity of $W[\theta,p]$    in Fig.\ref{fig:wignerplot_2}, where  $\langle L \rangle \simeq l+1/2$,  is more pronounced in  comparison with Fig.\ref{fig:wignerplot_1} in which case  $\langle L \rangle = l$. This  indicates that the   distribution $W[\theta,p]$  is, in particular, sensitive to the presence of   fractional expectation values of $\langle L \rangle$.

For  a general pure state  in terms of the first equality in Eq.(\ref{eq:wigner2024_3}),  we  now have  that
\begin{gather}  \label{eq:ourwigner_2}
    W[\theta,p] =  \frac{1}{2\pi }\sum_{n,m = -\infty }^{\infty}c(n)c^*(m) e^{\Dlarge{i(n-m)\theta }}  \int_{-\pi}^{\pi}\frac{ d\theta'}{2\pi} e^{\Dlarge{i(m+n-2p)\theta' }} ~ .
\end{gather}
This expression for $W[\theta,p]$ leads to the marginal distribution
\begin{gather} 
  W[p] =   \int_{-\pi}^{\pi}d\theta  W[\theta,p] =  \int_{-\pi}^{\pi} \frac{d\theta'}{2\pi} e^{\Dlarge{-i2p\theta' }}f(\theta') =  \sum_{n=-\infty}^{\infty} |c(n)|^2\mathrm{sinc}[2\pi(n-p)] ~ ,
   \label{eq:ourmarginal_2}
\end{gather}
where
\begin{gather} 
f(\theta) =    \sum_{n=-\infty}^{\infty} |c(n)|^2 e^{\Dlarge{i2n\theta}}  ~ ,
\end{gather}
and  $\mathrm{sinc}[x]=\sin[x]/x$.
We observe that for a general value of $p$ the marginal distribution $W[p]$    in Eq.(\ref{eq:ourmarginal_2}) is not necessarily positive definite.

\begin{figure}[htb]  
%
%\vspace{-0.5cm}
%
\centerline{\includegraphics[width=16cm,angle=0]{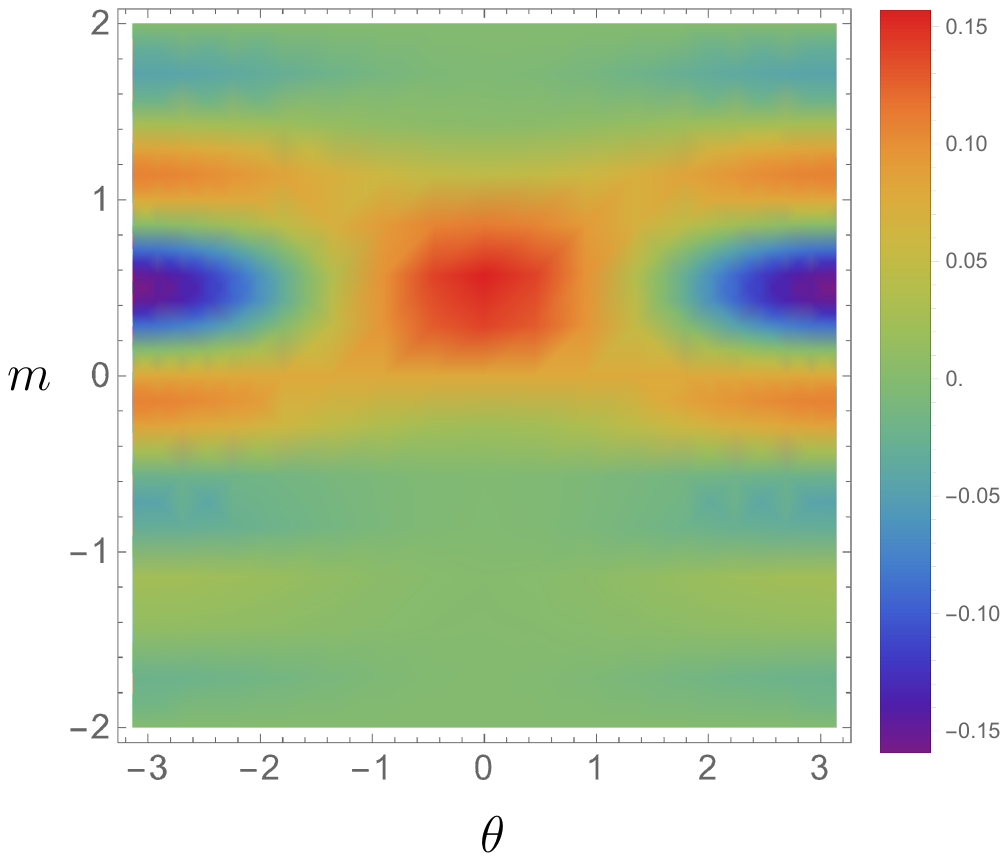}}
\vspace{-8.5cm}                 
\caption{The  distribution  $W[\theta,p]$ according to Eq.(\ref{eq:ourwigner_1}), with $p$ replaced by $m=p-l$   as in Fig.\ref{fig:wignerplot_2},  for $-\pi \le \theta \le \pi$ now  with 
$q=\exp(-1/2\lambda)=0.001$,  $\eps=1/2$,  and $\bar{l}= l +1/2$  in  the range $-2\le m \le 2$ for  the state $\psi(\theta)$ as in  Eq.(\ref{eq:wigner2024_3}).  The symmetry with respect to the  line  $m=0$ as in Fig.\ref{fig:wignerplot_1} is now not present.   For  half-integer values of $m$ the  distribution $W[\theta,m]$  can be negative  at $|\theta| = {\cal O}(\pi)$  and   the parameter  $p$  is fractional.
}
\label{fig:wignerplot_3}
\end{figure}

 If we, in particular,  consider the case  $\bar{l}=l + 1/2$ and  the state Eq.(\ref{eq:wigner2024_3})  in the limit of a small positive value of $\lambda$ as given by   Eq.(\ref{eq:lambda_zero})   we obtain
\begin{gather} 
W[\theta, p] = \frac{1}{4\pi} \large (\mathrm{sinc}[2\pi(l-p)] +2\mathrm{sinc}[2\pi(l-p+1/2)]\cos\theta    + \mathrm{sinc}[2\pi(l-p+1)] \large) \,  .
   \label{eq:ourhalfmarginal_2_0}
\end{gather} 
We observe that Eq.(\ref{eq:ourhalfmarginal_2_0})  turns out to be  in excellent  agreement with the numerical evaluation of $W[\theta, p]$ as presented in  Fig.\ref{fig:wignerplot_3}  for  $\lambda$-values  in the range $0 < \lambda \leq 0.1$.
Eq.(\ref{eq:ourhalfmarginal_2_0})  leads trivially  to the marginal distribution
\begin{gather} 
W[p] = \frac{1}{2} \large (\mathrm{sinc}[2\pi(l-p)] + \mathrm{sinc}[2\pi(l-p+1) ] \large) \,  ,
   \label{eq:ourhalfmarginal_2}
\end{gather} 
which is  not a positive definite function.   For the fractional angular momentum eigenstates $\psi_{l\gamma}(\theta)$ as defined in Eq.(\ref{eq:klauder_fractional}) one easily finds that 
\begin{gather} 
W[\theta, p]  = \frac{1}{2\pi} \mathrm{sinc}[2\pi(p - l -\gamma)]  \,  ,
   \label{eq:klauder_fractional2}
\end{gather} 
 which  now is $\theta$ independent and with a simple marginal distribution
\begin{gather} 
W[p] = \mathrm{sinc}[2\pi(p - l -\gamma)]  \,  ,
    \label{eq:klauder_fractional3}
\end{gather} 
which again  is  not a positive definite function.  In Section \ref{subsection-real_p} we will investigate in detail  how to calculate various moments of the distributions  given by Eqs.(\ref{eq:ourhalfmarginal_2_0})-(\ref{eq:klauder_fractional3}) if the variable $p$ can take any real value.

The negativity of   $W[\theta,p]$  at $p=l+1/2$   as in Fig.\ref{fig:wignerplot_2} can now be enhanced   by considering lower values of the parameter $q=\exp(-1/2\lambda)$. We then expect a dominant contribution for the state $\psi(\theta)$ in the form of Eq.(\ref{eq:lambda_zero}).   In a similar manner as in Fig.\ref{fig:wignerplot_2} and for the specific state  in Eq.(\ref{eq:wigner2024_3}),  we consider in Fig.\ref{fig:wignerplot_3}  the  distribution $W[\theta,p]$    as a function of $m=p - l$ instead as a function of $p$ with $\bar{l}=l + 1/2$  but  for $q = 0.001$.  For  other values  such that $q \leq 0.001$ the presence  of  negativity of the corresponding   distributions $W[\theta,m]$ remains essentially the same as  in Fig.\ref{fig:wignerplot_3}.      In the limit as $q \rightarrow 0$  we show in  Section \ref{subsection-real_p} that  the marginal distribution $W[p]$ leads  to the expectation value $\langle p \rangle =  \langle \psi| L |\psi\rangle/2 $, with the fractional expectation value $\langle \psi| L |\psi\rangle = l+1/2$ as obtained in Section \ref{sec:subsection_expectation_values}. 
 
In passing, we observe that it has been noticed (see, e.g.,  \Referes{Kastrup_2016,Kowalski_2021} and references cited therein) that  the  expression for $W[p]$ in Eq.(\ref{eq:ourmarginal_2})  is expressed  in  terms of  a Whittaker cardinal function $\mathrm{sinc}[\pi x]$  which is known  to  play an important role in signal analysis and important sampling theorems  in the reconstruction of functions from discrete sets of data (see, e.g., also \Referes{Whittaker_1915, Nyquist_1928, Shannon_1948, Whitney_1971,Stenger_1971,Gearhart_1990}).
%
%
%\newpage
%
%
\begin{center}
\subsection{ $W[\theta,p]$ when the range  of $p$ is all integers}
\label{subsection_integer_p}
\end{center}
%
%\vspace{-5mm}
%
%...\ref{subsectioninteger_p}
%
%\begin{comment}
%

If $p$ is an integer  the marginal distribution $W[p]$ in Eq.(\ref{eq:ourmarginal_2})  is, however,  in general positive and is   in accordance with Born's rule in quantum mechanics since Eq.(\ref{eq:ourwigner_2}) now  leads to  
\begin{gather} 
 \label{eq:ourwigner_10}
   W[p] =  \int_{-\pi}^{\pi}d\theta  W[\theta,p] =  |c(p)|^2  =|\int_{-\pi}^{\pi}d\theta \psi_p^*(\theta)\psi(\theta)|^2 ~ .
\end{gather}
It then follows that $\sum_{p= -\infty}^{\infty} W[p] =1$ as it should, which is the reason for  the overall  normalization factor which appears in Eq.(\ref{eq:ourwigner_1}).  The marginal distribution $W[p]$ in Eq.(\ref{eq:ourwigner_10}) can  be used to compute various expectation values of the observable $L$ reproducing the explicit results of \Refer{Skagerstam_2023}. 
%

%
%
% ------------   figure   (begin)    ---------------
%
%
% ------------   figure   (end)    ---------------
%
%
%\end{comment}

In order to obtain the marginal distribution $W[\theta]$ for the parameter $\theta$  when $p$    takes any integer value, we now invoke the reasonable physical requirement  that the Fourier expansion in Eq.(\ref{eq:wigner2024_3})  as well as in Eq.(\ref{eq:ourmarginal_2}) should make sense pointwise.  In terms of pure mathematics and for integers $p$ it is well-known that this is in general a delicate issue involving partial sums of Fourier series and the use of the rigorous mathematical Lebesgue method of integration as in \Refer{Carleson_1966}.  

 The expression for $W[\theta,p]$ in Eq.(\ref{eq:ourwigner_1}) for integers $p$  is, however, a well-defined and periodic in $\theta$.  In order to make sense of  the marginal distribution $W[\theta]=\sum_{p= -\infty}^{\infty} W[\theta,p]$ we  now make use of  the well established arithmetic Ces\`{a}ro sums  of  partial sums and  Fej\'{e}r summation methods of series (see, e.g., theorem $101$ in the first reference  in \Refer{Carslaw_1930}).  In practical  applications, like in signal analysis of experimental data, this means that we restrict any partial sum that enters in the analysis to be finite and therefore  summations and integrations can be interchanged. In the end we then formally  remove this restriction.  As is well-known, this procedure actually leads to pointwise convergence,  and it appears to be in line with actual experimental signal analysis.   For the convenience of the reader,   we  now explain in some detail  the method of Fej\'{e}r summation  of series which  takes boundary conditions into account.    
 
 We therefore first define  a partial sum expression  $\sigma_n(\theta)$  for  $W[\theta]$ by making use of  the second equivalent and convenient definition in Eq.(\ref{eq:ourwigner_1}), i.e.,
\begin{gather} 
 \sigma_n(\theta) = \sum_{p=-n}^{n} W[\theta,p] = \sum_{p=-n}^{n} \int_{-2\pi}^{2\pi}\frac{d\theta'}{4\pi}f(\theta,\theta'/2)
  e^{\Dlarge{-ip\theta' }} \nonumber \\ = \int_{-2\pi}^{2\pi}\frac{d\theta'}{4\pi} f(\theta,\theta'/2) \frac{\sin((n+1/2)\theta')}{\sin\theta'/2}\equiv   \int_{-2\pi}^{2\pi}\frac{d\theta'}{4\pi}  f(\theta,\theta'/2) D(n,\theta')  ~ ,
   \label{eq:ourwigner_11}
\end{gather}
where the finite function $f(\theta,\theta')$  in the range  $\theta, \theta' \in [-\pi,\pi]$ is  defined according to 
\begin{gather} 
 \label{eq:ourwigner_11_2}
f(\theta, \theta')= \psi(\theta+\theta')\psi^*(\theta -\theta') = \sum_{n,m}\Large(c(n)\frac{e^{\Dlarge{in\theta}}}{\sqrt{2\pi}}\Large) \Large(c^*(m)\frac{e^{\Dlarge{-im\theta}}}{\sqrt{2\pi}}\Large)e^{\Dlarge{i(m+n)\theta' }}~ ,
\end{gather}
which is $2\pi$-periodic in the variables $\theta$ and $\theta'$. Here we make use of    the definition of the  well-known periodic Dirichlet kernel $ D(n,\theta) = \sin((n+1/2)\theta)/\sin(\theta/2)$. A  periodic and positive Fej\'{e}r  kernel $F(M,\theta)$  defined by 
\begin{gather} 
F(M,\theta)=  \frac{1}{M}(\frac{\sin(M\theta/2)}{\sin(\theta/2)})^{\Dlarge{2}}   \, ,
 \label{eq:Our_Fejer}
\end{gather}
is then obtained by forming the arithmetic mean sum $\sigma(M,\theta) $ of the partial sum $\sigma_n(\theta)$  in Eq.(\ref{eq:ourwigner_11}), i.e.,  we define the Ces\`{a}ro sum
\begin{gather} 
\sigma(M,\theta)=  \frac{1}{M}  \sum_{k=0}^{M-1}  \sigma_k(\theta)=   \int_{-2\pi}^{2\pi}\frac{d\theta'}{4\pi}  f(\theta,\theta'/2) \frac{1}{M}(\frac{\sin(M\theta'/2)}{\sin(\theta'/2)})^2  \equiv  \int_{-2\pi}^{2\pi}\frac{d\theta'}{4\pi} f(\theta,\theta'/2) F(M,\theta') \, .
 \label{eq:ourwigner_12}
\end{gather}

In order to control the large $M$ singular behaviour of the   Fej\'{e}r kernel $F(M,\theta)$    at $\theta =0$ and $\theta = \pm 2\pi$  as used in Eq.(\ref{eq:ourwigner_12})
we split the range of integration into three  domains namely    $\theta' \in I_1=[-\pi,\pi]$,  $I_2=[-2\pi,-\pi]$, and $I_3=[\pi,2\pi] $.
Since the Fej\'{e}r  kernel  $F(M, \theta)$ is normalized in such a way that
\begin{gather} 
\label{eq:ourwigner_14}
 \int_{-\pi}^{\pi}\frac{d\theta}{2\pi}  F(M,\theta)=1~ ,
\end{gather} 
for all integers $M$ we can trivially write
\begin{gather} 
  \int_{I_1}\frac{d\theta'}{4\pi} f(\theta,\theta'/2) F(M,\theta') =  \frac{ f(\theta,0)}{2} + \int_{I_1}\frac{d\theta'}{4\pi} (f(\theta,\theta'/2) -f(\theta,0))F(M,\theta') \, .
 \label{eq:ourwigner_13}
\end{gather}
The range of integration over $\theta' \in I_1$ in Eq.(\ref{eq:ourwigner_13}) is then  divided into $-\delta <\theta' <\delta$, and  $\delta \leq |\theta'| \leq \pi$,  where $\delta >0$ is assumed to be sufficiently small.  In the range $-\delta <\theta' <\delta$  the contribution to the integration can then be made  arbitrarily small due to the continuity properties of $f(\theta, \theta')$. In the range $\delta\leq|\theta'|\leq\pi$ we can, on the other hand,  make use of the bound $F(M,\theta') \leq 1/ M\sin^2(\delta/2)$ and make the contribution arbitrarily small in the large $M$-limit.  For sufficiently large values of $M$  the  integral in Eq.(\ref{eq:ourwigner_13})  can then be made arbitrarily  small in both ranges of integration  and the only contributing term  is $f(\theta,0)/2$ which corresponds to  Fej\'{e}rs theorem.  

In a similar manner we consider
\begin{gather} 
  \int_{I_2}\frac{d\theta'}{4\pi} f(\theta,\theta'/2) F(M,\theta') =  \frac{ f(\theta,-\pi)}{4} + \int_{0}^{\pi}\frac{d\theta'}{4\pi} (f(\theta,\theta'/2 - \pi) -f(\theta,-\pi))F(M,\theta') \, ,
 \label{eq:ourwigner_13_2}
\end{gather}
after a shift of the variable of integration according to $\theta' \rightarrow \theta'+2\pi$ and using the periodicity $F(M,\theta'-2\pi)= F(M,\theta')$. The last integral in Eq.(\ref{eq:ourwigner_13_2}) can then be treated  as in Eq.(\ref{eq:ourwigner_13}).  For $\theta' \in I_3 $ we proceed in a similar manner using a shift of integration variable according to $\theta' \rightarrow \theta'-2\pi$. 
In the limit of large  $M$ the marginal distribution $W[\theta]=\sum_{p= -\infty}^{\infty} W[\theta,p]$ to be obtained from Eq.(\ref{eq:ourwigner_1}) for integers $p$  is now  defined to be given by the expression $\lim_{ M\rightarrow \infty}\sigma(M,\theta)$  and therefore leads to
\begin{gather} 
W[\theta]  = \frac{1}{2}\left( |\psi(\theta)|^2 +  \frac{1}{2}( |\psi(\theta - \pi)|^2 + |\psi(\theta + \pi)|^2 )  \right) =   \frac{1}{2}\left( |\psi(\theta)|^2 +  |\psi(\theta + \pi)|^2   \right) \, ,
 \label{eq:ourwigner_16}
\end{gather}
where we make use of the $2\pi$-periodicity of the state $\psi(\theta)$  as defined in  Eq.(\ref{eq:wigner2024_3}). The marginal distribution $W[\theta]$ so defined is  correctly normalized after an integration over the angle $\theta$ in the range  $\theta \in  [-\pi, \pi]$. 

% ------------   figure   (begin)    ---------------
%
\begin{figure}[htb]  
\vspace{-1cm}
%
%
%\vspace{-13.0cm}                 
\centerline{\includegraphics[width=16cm,angle=0]{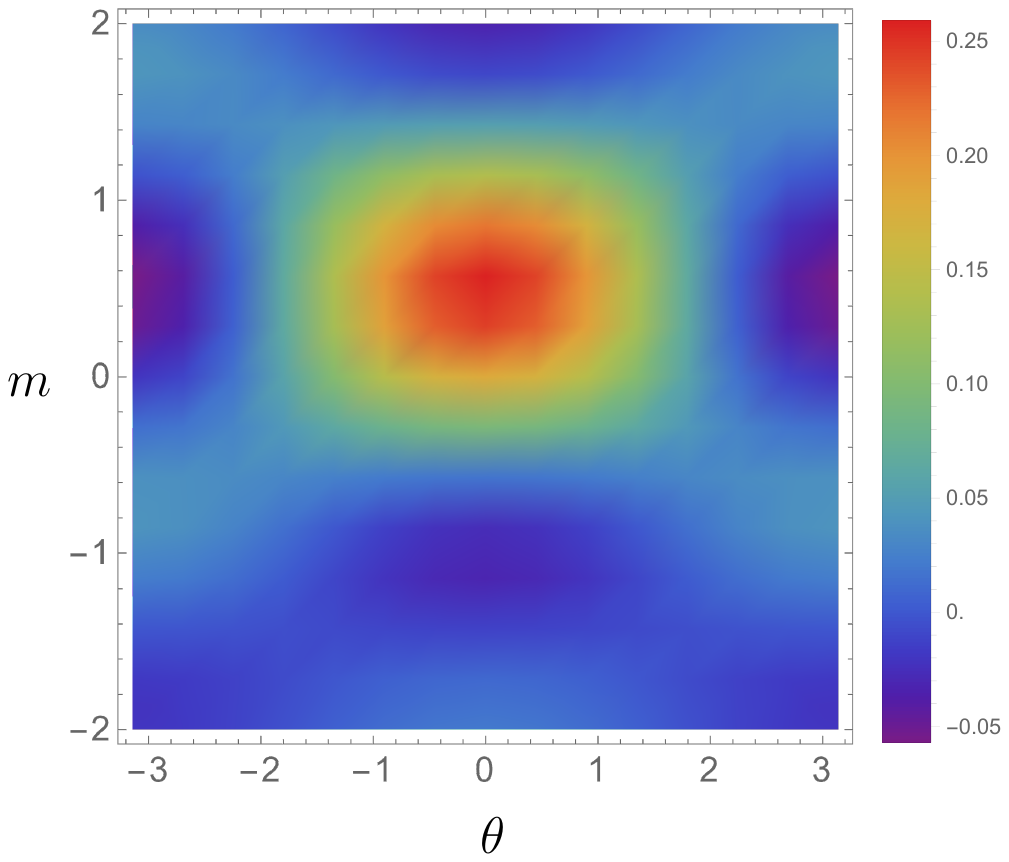}}
 \vspace{-9.0cm}
\caption{The distribution  $W_{1/2}[\theta,p]$ according to Eq.(\ref{eq:wigner_kastrup}), with $p$ replaced by $m=p-l$   as in Fig.\ref{fig:wignerplot_3},   for $-\pi \le \theta \le \pi$   with $q=0.001$,  $\bar{l}= l +\eps$, and $\eps =1/2$ for  the state $\psi(\theta)$ according to Eq.(\ref{eq:wigner2024_3})  in the range $-2\leq m \leq 2$. As in Fig.\ref{fig:wignerplot_3} the symmetry with respect to the line with $m=0$ is absent.  Regions of negativity for  $W_{1/2}[\theta,m]$ are in general not the same as  in  Fig.\ref{fig:wignerplot_3} for the   distributions $W[\theta,m]$ with the same quantum state $\psi(\theta)$ considered. 
}
\label{fig:wignerplot_4}
\end{figure}

A less precise and formal procedure in order to obtain Eq.(\ref{eq:ourwigner_16}) is to make use of the Dirac comb expression
 (see, e.g., \Refer{Giraud_2016})  in Eq.(\ref{eq:ourwigner_2}), i.e., 
\begin{gather} 
 \frac{1}{2\pi}  \sum_{p=-\infty}^{\infty}  e^{\Dlarge{-ip\theta}} =\sum_{p=-\infty}^{\infty} \delta(\theta - 2\pi p)  \, ,
 \label{eq:ourwigner_17}
\end{gather}
in terms of a periodic Dirac $\delta(\theta)$-function with  formally a half contribution at the boundary of integration  $\theta = \pm 2\pi$.

We now observe the following alternative expression for the Fej\'{e}r kernel $F(M,\theta)$ as used in Eq.(\ref{eq:ourwigner_12}) namely
\begin{gather} 
F(M,\theta) = \sum_{p=-M}^{M}  (1- |\frac{p}{M}|) e^{\Dlarge{-ip\theta}}  \, ,
 \label{eq:Fejer_identity}
\end{gather}
which is straightforward to verify (in this context see, e.g., \Refer{Grafakos_2014}). If the integers $p$ are interpreted as the discrete spectrum of the observable $L$,  the bracket in Eq.(\ref{eq:Fejer_identity}) can be regarded as an effective cut-off of this spectrum and makes an  otherwise  infinite sum over $p$  without this bracket finite.  If such a cut-off is considered in the data analysis of an experimental situation,   we may again be led to   Eq.(\ref{eq:ourwigner_16})  in the large $M$-limit. In the next section we will consider an  extension of  Eq.(\ref{eq:Fejer_identity}) when the summation over the integer $p$ is replaced by suitable  integral over  the parameter $p$.  
%
%
%\newpage
%
%
\begin{center}
\subsection{ $W[\theta,p]$ when the range  of $p$ is all real numbers \label{subsection-real_p}}
\end{center}
%
%\vspace{-5mm}

The distribution $W[\theta,p]$ in Eq.(\ref{eq:ourwigner_1}) can now  be regarded as  a well-defined extension of a function  not only for integer values of  $p$  but also for any real number $p$.   For real numbers $p$ the corresponding marginal distribution  $W[p]$ is,   however,  then  not a necessarily a positive  function of $p$  and can  therefore not in general be interpreted as a probability distribution. The physical meaning of the marginal  $W[p]$ for all real numbers is, furthermore, not entirely clear since the measure of rational numbers in the set of real numbers is zero (see, e.g., \cite{Born_1955}).

  In this work we  focus our attention on    fractional values of the parameter $p$  but here we  nevertheless find it useful to recall the following known extension of the Ces\'{a}ro summability methods for integers as used in Section \ref{subsection_integer_p} to  ill-defined integrals over $p$.  Therefore Eq.(\ref{eq:ourwigner_12}) is  extended to the large $\lambda$-limit of the following expression
\begin{gather} 
 \sigma(M,\theta)\rightarrow  \sigma(\lambda, \theta)  =  \int_{-2\pi}^{2\pi}\frac{d\theta'}{4\pi}
  \int_{-\lambda}^{\lambda} dp (1- |\frac{p}{\lambda}|)f(\theta,\theta'/2)e^{\Dlarge{-ip\theta' }}   ~ ,
 \label{eq:cintegral0}
\end{gather}
to be used as  a well-defined definition of $W[\theta]=\lim_{\lambda \rightarrow \infty} \sigma(\lambda, \theta)$. Here 
 we make use of the following  extension of Eq.(\ref{eq:Fejer_identity}) to Ces\'{a}ro summability of integrals  which is easy to verify namely
\begin{gather} 
\frac{1}{\lambda}\int_{0}^{\lambda} dn \int_{-n}^{n}dps(p) 
= \int_{-\lambda}^{\lambda} dp (1- |\frac{p}{\lambda}|)s(p)   \, ,
 \label{eq:Titchmarshidentity}
\end{gather}
for a function $s(p)$ (see, e.g., Chapter I and  subsection 1.15 in Ref.\cite{Titchmarsh_1948} for a slightly  simpler version of Eq.(\ref{eq:Titchmarshidentity})). We will make use of the right-hand expression of Eq.(\ref{eq:Titchmarshidentity}) in actual computations.    Below we, however, verify that  the use of Eq.(\ref{eq:Titchmarshidentity})  will not always give rise to
well-defined  expressions in the limit of  large $\lambda$. If required  we can  then improve the large $\lambda$ behaviour  by considering iterative replacements like $s(p) \rightarrow (1- |p/\lambda|)s(p)$ in Eq.(\ref{eq:Titchmarshidentity}). 

With the choice  $s(p) = \exp(-ip\theta')$ we  now obtain the result

\begin{gather} 
 \label{eq:cintegral1}
  \sigma(\lambda, \theta)  =  \int_{-2\pi}^{2\pi}\frac{d\theta'}{4\pi}
f(\theta,\theta'/2)K_{\lambda}(\theta')  =  \int_{-\lambda\pi}^{\lambda\pi}\frac{dx}{2\pi}
f(\theta,x/\lambda)(\frac{\sin x}{x})^2 ~ ,
\end{gather}
in terms of the function $f(\theta,\theta')$ in Eq.(\ref{eq:ourwigner_11_2}),  and
where we have defined the positive kernel $K_{\lambda}(\theta)$ by 
\begin{gather} 
K_{\lambda}(\theta) = \frac{1}{\lambda}  (\frac{\sin(\lambda \theta/2)}{\theta/2})^2~ .
   \label{eq:ourKernel}
\end{gather}
We observe that the kernel $K_{\lambda}(\theta)$ is a regular function at the boundary points $\theta = \pm 2\pi$ and is normalized according to 
\begin{gather} 
 \lim _{\lambda \rightarrow \infty}\int_{-2\pi}^{2\pi}\frac{d\theta}{4\pi} K_{\lambda}(\theta) = \frac{1}{2} \, .
  % \label{eq:ourwigner_11}
\end{gather}

The method as considered  in the evaluation  of the large $M$-limit  of  $\sigma(M,\theta)$  in Eq.(\ref{eq:ourwigner_12})  can now be applied in order to evaluate the large $\lambda$-limit of $\sigma(\lambda, \theta)$ in Eq.(\ref{eq:cintegral1}). The kernel $K_{\lambda}(\theta)$ is then  such that only an interval like $I_1$ as used in 
Eq.(\ref{eq:ourwigner_13})  contributes   in the large $\lambda$-limit. Alternatively  one can consider the large $\lambda$-limit of the last expression in Eq.(\ref{eq:cintegral1})  and one  finds that
\begin{gather} 
 W[\theta] = \lim_{\lambda \rightarrow \infty} \sigma(\lambda, \theta) = \frac{1}{2} |\psi(\theta)|^2~ .
  \label{eq:our_theta_normalization}
\end{gather}

In order to compute moments $\langle p^{\alpha} \rangle$ of the variable $p$ for integers $\alpha \geq 0$ using the distribution $W[p]$, we proceed as in the discussion in Section \ref{subsection_integer_p} and first make   the sum in Eq.(\ref{eq:ourmarginal_2}) finite by a restriction $|n| \leq M$ for some sufficiently large integer $M$.
For integers $\alpha, \beta \geq 0$, and in  terms of the Ces\'{a}ro summability for integrals in Eq.(\ref{eq:Titchmarshidentity}), we then make sense of the formal expression for  $\langle p^{\alpha} \rangle$  by considering the general replacement  
\begin{gather} 
\label{eq:ourmarginal_alpha}
 \langle p^{\alpha} \rangle \equiv  \int_{-\infty}^{\infty}dp  p^{\alpha}  W[p] \rightarrow \sum_{|n|\leq M}|c(n)| ^2 \lim_{\lambda \rightarrow \infty}  \int_{-\lambda}^{\lambda} dp p^{\alpha}  (1- |\frac{p}{\lambda}|)^{\beta}\mathrm{sinc}[2\pi(p-n))] ~ .
\end{gather}
For the marginal distribution as defined  in $W[p]$ in Eq.(\ref{eq:ourmarginal_alpha})  with $\alpha =\beta=0$,   it is then straightforward to, e.g., verify that in the large $M$-limit
\begin{gather} 
\label{eq:our_p_normalization}
    \int_{-\infty}^{\infty}dp  W[p] = \frac{1}{2} ~ ,
\end{gather}
which is consistent with the normalization of the marginal distribution $W[\theta]$ in Eq.(\ref{eq:our_theta_normalization}).

It is now a remarkable property of the  large $\lambda$-limit in  Eq.(\ref{eq:ourmarginal_alpha}) that the result  Eq.(\ref{eq:our_p_normalization}) for $\alpha = 0$ is valid for all $\beta \geq 0$ as we have verified by computing the relevant integrals in an exact manner (for a related discussion see Chapter I and  subsection 1.15 in Ref.\cite{Titchmarsh_1948}).  In the case with $\alpha= 1$ and $\beta=0$ one, however,   finds that Eq.(\ref{eq:ourmarginal_alpha}) is ill-defined. With $\alpha=\beta = 1$ the Ces\'{a}ro summability as used in Eq.(\ref{eq:ourmarginal_alpha}) turns out to be  well-defined in the large $\lambda$-limit and  is independent of all integers $\beta \geq 1$ if $\alpha =1$.   For $\beta =1$ the moment $\langle p^{2} \rangle$ is, however, then ill-defined and can be well-defined by making use of the iterative procedure mentioned in connection with the Ces\'{a}ro summability procedure  defined in Eq.(\ref{eq:Titchmarshidentity}).  With $\alpha=\beta=2$ it turns out  that $\langle p^{2} \rangle$ is well-defined and independent of integers  $\beta \geq 2$ if $\alpha =2$.  In general these non-trivial results  can now be extended in such a way  that    for any normalized state $\psi(\theta)$ and in  the large $M$-limit in Eq.(\ref{eq:ourmarginal_alpha})
 \begin{gather} 
 %\label{eq:ourwigner_10_2}
 \langle p^{\alpha} \rangle =  \frac{1}{2 }\sum_{n=-\infty}^{\infty} n^{\alpha} |c(n)| ^2  =  \frac{1}{2} \langle  L^{\alpha} \rangle ~ ,
\end{gather}
for integers $\alpha$ and $\beta$ such that $\beta \geq \alpha$ where, of course, $\langle  L^{\alpha} \rangle = \langle \psi| L^{\alpha} |\psi\rangle$ is the expectation value  in quantum mechanics as discussed in Section \ref{sec:subsection_expectation_values}.
%
%%using similar and straightforward  methods as, e.g., discussed in Ref.\cite{Kowalski_2021}.
%
%
\newpage
\begin{center}
\section{THE QUASI-PROBABILITY DISTRIBUTION ${\bm{W_{1/2}[\theta,p]}}$
\label{section_polish_W}}
\end{center}
\begin{center}
\subsection{ A Finite Domain for the Parameter  $\theta$
\label{subsection_polish_Finite_Domain}}
\end{center}

 Another definition of a quasi-probability  distribution considered  in great theoretical detail  in the literature  (see, e.g., \Referes{Kastrup_2016,Kowalski_2021} and references cited therein) is 
\begin{gather}  \label{eq:wigner_kastrup}
   W_{1/2}[\theta,p] = \int_{-\pi}^{\pi} \frac{d\theta'}{2\pi} e^{\Dlarge{-ip\theta' }} \psi(\theta+\theta'/2) \psi^*(\theta -\theta'/2) ~ ,
\end{gather}
and the Cauchy-Schwarz inequality now implies  the bound  $|W_{1/2}[\theta,p]| \leq 1/\pi$.  It then follows that 
Eq.(\ref{eq:ourmarginal_2}) is replaced by 
\begin{gather} 
  W_{1/2}[p] =   \int_{-\pi}^{\pi}d\theta  W_{1/2}[\theta,p]  =  \sum_{n=-\infty}^{\infty} |c(n)|^2\mathrm{sinc}[\pi(n-p)] ~ ,
   \label{eq:halfmarginal_2}
\end{gather}
which for integer values of $p$ is in accordance with Born's rule in quantum mechanics  as in Eq.(\ref{eq:ourwigner_10}). Proceeding as in Section \ref{subsection_Finite_Domain} for the quasi-distribution $W[\theta,p]$,   we can regard $W_{1/2}[\theta,p]$ in Eq.(\ref{eq:halfmarginal_2})  to be a well-defined function also for any real number $p$.  We therefore again conclude that the marginal distribution  $W_{1/2}[p]$ is  not always  a positive function of $p$, i.e., it cannot in general be interpreted as a probability distribution. Furthermore one obtains equations similar to Eqs.(\ref{eq:ourhalfmarginal_2_0})-(\ref{eq:klauder_fractional3})  by a simple change of factors of $\pi$. If we, e.g.,  consider the state $\psi(\theta)$  in Eq.(\ref{eq:wigner2024_3}) with  $\bar{l}=l + 1/2$ and the  limit of a small positive value of $\lambda$, i.e.,  the state as  in Eq.(\ref{eq:lambda_zero}),   we now  have that Eq.(\ref{eq:ourhalfmarginal_2_0}) is replaced by 
\begin{gather} 
W_{1/2}[\theta, p] = \frac{1}{2\pi} \large (\mathrm{sinc}[\pi(l-p)] +2\mathrm{sinc}[\pi(l-p+1/2)]\cos\theta    + \mathrm{sinc}[\pi(l-p+1)] \large) \,  .
 \label{eq:klauder_halfmarginal_2_0}
\end{gather} 

As for a comparison  with Fig.\ref{fig:wignerplot_3}  we illustrate in  Fig.\ref{fig:wignerplot_4}  the  distribution $W_{1/2}[\theta,p]$  for the state  in Eq.(\ref{eq:wigner2024_3}) with $q= \exp(-1/2\lambda)=0.001$  as a function of $m=p - l$ instead as a function of $p$ in the case with $\bar{l}=l + 1/2$.  For  other values  such that $q\leq 0.001$ the presence  of  negativity of the corresponding  distributions $W_{1/2}[\theta,m]$   remain essentially the same as illustrated in Fig.\ref{fig:wignerplot_4}.      In the limit as $q \rightarrow 0$ we obtain the half-integer  expectation value $\langle L \rangle = l+1/2$.
In this limit the distribution $W_{1/2}[\theta,m]$  takes the form as in Eq.(\ref{eq:klauder_halfmarginal_2_0}) and can   then clearly be  negative  when $p$ is fractional.  The negativity of $W[\theta,p]$  for fractional values of $p$ is clearly more localized in  Fig.\ref{fig:wignerplot_3} in comparison with what we exhibit for $W_{1/2}[\theta,p]$ in  Fig.\ref{fig:wignerplot_4}. 

 Similar to the use of Eq.(\ref{eq:ourhalfmarginal_2_0})  for the  Fig.\ref{fig:wignerplot_3},  we also find that the use of the  analytical expression Eq.(\ref{eq:klauder_halfmarginal_2_0}) is in good agreement with  the numerical evaluations in Fig.\ref{fig:wignerplot_4} for a broad range of values of the parameter $\lambda$ in the range 
$< \lambda \leq 0.1$.
%
%\vspace{0.5cm}
%
\newpage
\begin{center}
\subsection{ $W_{1/2}[\theta,p]$ when the range  of $p$ is all integers}
\end{center}
%
%\vspace{-5mm}
%
%

In the case of  integer values of the parameter $p$  we have verified above that  $W_{1/2}[p] =|c(p)|^2$.
The marginal distribution $W_{1/2}[\theta]= \sum_{p=-\infty}^{\infty} W_{1/2}[\theta,p]$   can be derived   by again making use of the  Fej\'{e}r summation methods as discussed in Section \ref{subsection_integer_p} for the quasi-probability  distribution $W[\theta,p]$. In the second equivalent form of the   distribution $W[\theta,p]$ in Eq.(\ref{eq:ourwigner_1}) the range of integration  is   $\theta' \in [-2\pi,2\pi]$.  We therefore first  observe that  for the  distribution $W_{1/2}[\theta,p]$ the corresponding  range of integration in Eq.(\ref{eq:wigner_kastrup})  is instead given by $\theta' \in [-\pi,\pi]$. With the range of $|\theta \pm \theta'/2| \le 3\pi/2$ in the states $\psi(\theta \pm \theta'/2)$  in Eq.(\ref{eq:wigner_kastrup}),  it  therefore immediately follows from the analysis in Section \ref{subsection_integer_p} and  Eq.(\ref{eq:ourwigner_13}),  or  by making use of the  less precise periodic $ \delta(\theta)$  function in Eq.(\ref{eq:ourwigner_17}),   that   the  marginal distribution $W_{1/2}[\theta]$ is now given by 
\begin{gather}
W_{1/2}[\theta] =|\psi(\theta)|^2  \, ,
 \label{eq:theta_halfmarginal_2}
\end{gather}
to be compared to Eq.(\ref{eq:ourwigner_16}) in Section \ref{subsection_integer_p}.  We therefore conclude  that marginal distribution $W_{1/2}[\theta]$ implies the correct overall normalization of  $W_{1/2}[\theta,p]$ according to 
\begin{gather}
\lim_{n\rightarrow \infty} \sum_{p=-n}^{n} \int_{-\pi}^{\pi}d\theta W_{1/2}[\theta,p] = \int_{-\pi}^{\pi}d\theta  W_{1/2}[\theta] =1 \, .
\end{gather}
%
%
%
%
%\vspace{1mm}
%
%
%\vspace{1cm}
%

%
%\newpage
%
%
\begin{center}
\subsection{$W_{1/2}[\theta,p]$ when the range  of $p$ is all real numbers}
\end{center}
%
%\vspace{1mm}

For integers $p$ we have verified above  that even if the distribution  $W_{1/2}[p]$ is in accordance with Born's rule  in quantum mechanics it is, however,  not in general a positive function for all real numbers $p$.  
With the same conditions as in the derivation of Eq.(\ref{eq:ourhalfmarginal_2}) one, e.g.,  finds that
\begin{gather} 
W_{1/2}[p] = \frac{1}{2} \large (\mathrm{sinc}[\pi(l-p) + \mathrm{sinc}[\pi(l-p+1) ] \large) \,  ,
 %  \label{eq:halfmarginal_2}
\end{gather} 
which is not in general a positive definite function as function of $l-p$.
It is therefore not clear in what sense the distribution  $W_{1/2}[p]$ should be normalized. We therefore proceed as in Section  \ref{subsection-real_p} and make use of the Ces\'{a}ro summability of integrals.  

In case of the  distribution $W_{1/2}[\theta]$ we therefore  define it  in terms of the limit $\lim_{\lambda \rightarrow \infty}\sigma_{1/2}(\lambda,\theta)$,   where now $\sigma(\lambda, \theta)$ in Eq.(\ref{eq:cintegral0}) is replaced by the well-defined  expression 
\begin{gather} 
 \sigma_{1/2}(\lambda, \theta)  =  \int_{-\pi}^{\pi}\frac{d\theta'}{2\pi}
  \int_{-\lambda}^{\lambda} dp (1- |\frac{p}{\lambda}|)f(\theta,\theta'/2)e^{\Dlarge{-ip\theta' }}  =
  \int_{-\pi}^{\pi}\frac{d\theta'}{2\pi}
f(\theta,\theta'/2)K_{\lambda}(\theta')     ~ .
 \label{eq:cintegral2}
\end{gather}
Proceeding as in Section \ref{subsection-real_p}  it therefore follows that  the distribution $W_{1/2}[\theta]$ is   given by the same expression as for integers in Eq.(\ref{eq:theta_halfmarginal_2}), i.e., 
\begin{gather} 
 W_{1/2}[\theta] =  \int_{-\infty}^{\infty}dp  W_{1/2}[\theta,p] = \lim_{\lambda \rightarrow \infty}\sigma_{1/2}(\lambda,\theta)= |\psi(\theta)|^2~ ,
  \label{eq:cintegral3}
\end{gather}
 to be compared to Eq.(\ref{eq:our_theta_normalization}) for the distribution $W[\theta]$. The procedure above to find $W_{1/2}[\theta]$ justifies  the less precise and formal method  as used in Ref.\cite{Kowalski_2021}  to obtain the same result. 

If $p$ is any real number we can make use of   Eq.(\ref{eq:halfmarginal_2})  and directly verify that the normalization in Eq.(\ref{eq:our_p_normalization}) is
\begin{gather} 
 \label{eq:halfwigner_10_2}
    \int_{-\infty}^{\infty}dp  W_{1/2}[p] = 1 ~ ,
\end{gather}
 to be compared with the normalization condition Eq.(\ref{eq:our_p_normalization}) for the distribution $W[p]$.
 
Expectation values of powers of the parameter $p$ using the distribution $W_{1/2}[p]$ leads to ill-defined expressions.
Proceeding again as in Section \ref{subsection-real_p} we therefore consider as in Eq.(\ref{eq:ourmarginal_alpha}) the general expression
\begin{gather} 
%\label{eq:ourmarginal_alpha}
 \langle p^{\alpha} \rangle \equiv  \int_{-\infty}^{\infty}dp  p^{\alpha}  W_{1/2}[p] \rightarrow \sum_{|n|\leq M}|c(n)| ^2 \lim_{\lambda \rightarrow \infty}  \int_{-\lambda}^{\lambda} dp p^{\alpha}  (1- |\frac{p}{\lambda}|)^{\beta}\mathrm{sinc}[\pi(p-n))] ~ ,
\end{gather}
for integers $\alpha, \beta \geq 0$. One then finds that   $\langle p^{\alpha} \rangle=\langle \psi|  L^{\alpha} |\psi\rangle$ in the large $M$-limit for all integers $\alpha$ and $\beta$ such that $\beta \geq \alpha$.

%\begin{comment}
% ------------   figure   (begin)    ---------------
%
\begin{figure}[htb]  
%
%\vspace{-0.5cm}
%
%
%\vspace{-0.5cm}                 
\centerline{\includegraphics[width=17cm,angle=0]{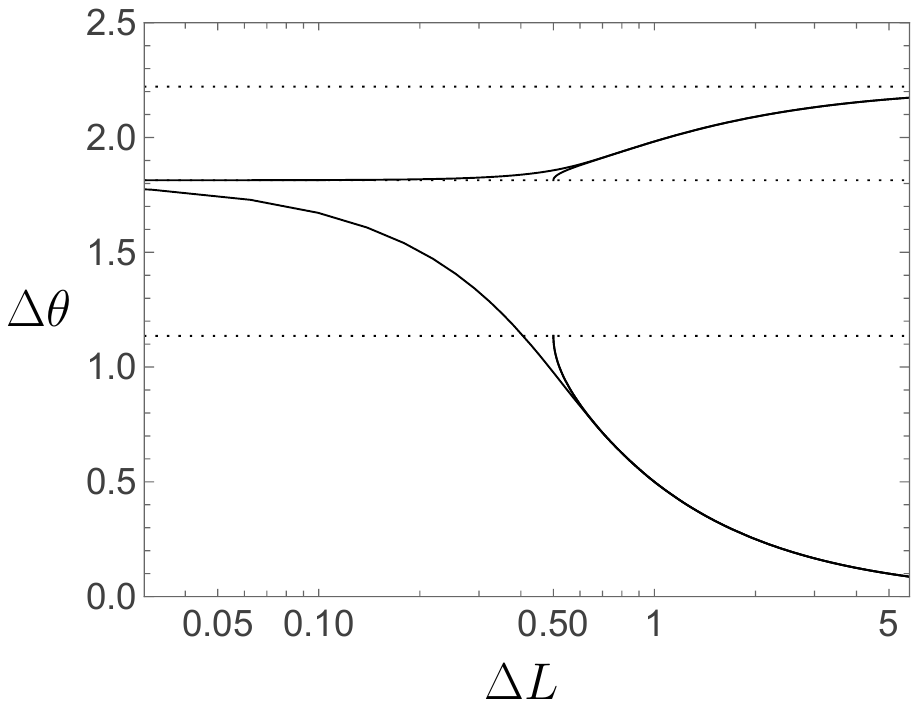}}
\vspace{-12cm}
\caption{The  lower and  upper sold curves correspond to  $\Delta\theta$ as  a functions of  $\Delta L$ in a log-scale obtained using  $p_1 (\theta)$  or $p_2 (\theta)$, respectively,  defined in the main text
% for  the state   in Eq.(\ref{eq:wigner2024_3}) 
with  $\bar{l} = l + \eps$ and  $\bar{\theta}=0$.  $\Delta\theta$  and $\Delta L$  are then independent of the integer $l$. The lower and upper solid curves that  end at $\Delta L = 1/2$ correspond to  the   rational value $\eps =1/2$. The lower and upper solid curves that coalesce at $\Delta L = 0$  and $\Delta \theta = \pi /\sqrt{3}$  correspond to   $\eps \neq 1/2$.
The two upper  horizontal dotted lines correspond to the bounds $\pi/\sqrt{2}$  and $\pi/\sqrt{3}$, respectively. The lower dotted line corresponds to   the upper bound $\Delta\theta=\sqrt{\pi^2/3 -2}$  with the uncertainty $\Delta L =1/2$  when $\eps=1/2$  in accordance with predictions of  \Refer{{Skagerstam_2023}}. 
}
\label{fig:wignerplot_5}
\end{figure}
%\end{comment}
%

%\begin{comment}
%
\begin{center}
\section{MEASUREMENT OF UNCERTAINTIES
\label{section_our_uncertainties}}
\end{center}
\begin{center}
\subsection{The Probability Distributions $p_1(\theta)$ and  $p_2(\theta)$
\label{subsection_measurements_1}}
\end{center}

We have observed that  the state-dependent uncertainties $\Delta \theta$ and $\Delta L$ do not depend on the parameter $l$ in the quantum state Eq.(\ref{eq:wigner2024_3}). Without resorting to   general properties of quasi-probability  distributions like $W[\theta,p]$ or $W_{1/2}[\theta,p]$, we now raise the question to what extent  such state-dependent observables can be used to directly observe  signs of more peculiar properties of  states  in the form of Eq.(\ref{eq:wigner2024_3}) in particular when the parameter $\bar{l}$ is a half-integer $\bar{l}=l+1/2$.

 In the spirit of the theoretical as well as the experimental work considered in  \Refer{Padgett_2004},  we   will now verify that this is indeed possible. As discussed in more detail in \Refer{Skagerstam_2023} for the state in Eq.(\ref{eq:wigner2024_3}),  the procedure in  \Refer{Padgett_2004} is based on the   elimination of  the dependence of the parameter $\lambda$ in terms of the state-dependent observables $\Delta \theta$ and $\Delta L$. One then considers the statistics of  many prepared states of the same form but only with different  values of the parameter $\lambda$ in order to obtain data for  the observables  $\Delta \theta$ and $\Delta L$.

 In this manner one can then, as we verify below,  describe  some special features of states in  the form of Eq.(\ref{eq:wigner2024_3})  by making use of only the knowledge of a measurable    probability  distribution  defined by 
\begin{gather}
 p_1(\theta) \equiv |\psi(\theta)|^2 \, .  
 \label{eq:half_probability}
 \end{gather}
  In a similar manner we also define the probability distribution $p_2(\theta)$ according to 
\begin{gather}
p_2(\theta) \equiv  \frac{1}{2} \left( p_1(\theta) +p_1(\theta + \pi) \right )\, .  
 \label{eq:whole_probability}
 \end{gather}
 The probability distributions $p_2(\theta)$ and $p_1(\theta)$ have, of course, a correspondence to the well-defined 
 marginal probability distributions $W[\theta]$ and  $W_{1/2}[\theta]$ as obtained from the quasi-probability distributions  $W[\theta,p]$ and $W_{1/2}[\theta,p]$  for integer values of the parameter $p$ as discussed in Sections \ref{section_ourW} and \ref{section_polish_W}, respectively.
We leave it open to what extent the  probability distributions $p_1(\theta)$ and/or $p_2(\theta)$ correspond to experimentally accessible data  but will show that they give rise to quite different predictions. 

 In  Fig.\ref{fig:wignerplot_5}  we display $\Delta \theta$ as a function of $\Delta L$  by making use  of the probability distribution  $p_1(\theta)$ or  $p_2(\theta)$ for  quantum states of the form as in Eq.(\ref{eq:wigner2024_3}) with a fixed fractional value   $\bar{l} = l + 1/2$,  where $l$ is an integer.  For each  value of  the parameter $\lambda$  one then obtains  a corresponding  pair of uncertainties $\Delta \theta$ and $\Delta L$.  The limits of small or large values of the parameter $\lambda$  can now be analysed in detail in order to explain characteristic  features  of Fig.\ref{fig:wignerplot_5}  obtained from a   numerical evaluation using the exact state  in Eq.(\ref{eq:wigner2024_3}).
 
 \begin{center}
\subsection{$p_1(\theta)$ and Expansions  of the Parameter $\lambda$
\label{subsection_measurements_1}}
\end{center}

 As discussed  in Section \ref{section_our_state},  the use of Eq.(\ref{eq:wigner2024_3})  with $\bar{l} = l + 1/2$ and in the limit of small $\lambda$  leads to the   superposition of states  in Eq.(\ref{eq:lambda_zero}) with expectation values $\langle L \rangle =  l + 1/2$ and $\Delta L =1/2$.  For $\bar{l} = l + \eps$ and  $0 \leq \eps <1$,  the limit of small $\lambda$ leads to angular momentum eigenstates $\psi_l(\theta)$ or  $\psi_{l+1}(\theta)$ with $\Delta \theta = \pi /\sqrt{3}$, and $\Delta L =0$ if $\eps \neq 1/2$.
 %
 %
% As observed in  \Refer{Skagerstam_2023} that

In  the small $\lambda$-limit  of the quantum state $\psi(\theta)$  in  Eq.(\ref{eq:wigner2024_3})   
it is now convenient to make use of Eq.(\ref{eq:alternative_form}) in order to find $p_1(\theta)$  by expanding  the theta functions $\vartheta_3 [\theta/2-i\eps/2\lambda,q]$  and $\vartheta_3 [-i\eps/\lambda,q^2]$   in powers of $q=\exp(-1/2\lambda)$. With $\bar{\theta}=0$, the definition  in Eq.(\ref{eq:two_phase_6}) then leads to the expressions 
\begin{gather} 
\vartheta_3 [\theta/2-i\eps/2\lambda,q] =  1 + e^{\displaystyle{\epsilon/\lambda}}e^{\displaystyle{i\theta}}q +
e^{\displaystyle{-\epsilon/\lambda}}e^{\displaystyle{-i\theta}}q  +  {\cal O}(q^4) \, ,
 \label{eq:small_q1_form}
\end{gather}
and
\begin{gather} 
\vartheta_3 [-i\eps/\lambda,q^2] =  1 +e^{\displaystyle{2\epsilon/\lambda}}q^2 +
e^{\displaystyle{-2\epsilon/\lambda}}q^2  +    {\cal O}(q^8)  \, .
\label{eq:small_qnorm_form}
\end{gather}
It is now clear from Eqs.(\ref{eq:small_q1_form}) and (\ref{eq:small_qnorm_form}) that the rational number $\epsilon =1/2$ plays a special role as has already been remarked above.
%%in Section \ref{sec:subsection_expectation_values}. 
In the limit of a sufficiently small $\lambda$  with $\eps < 1/2$ it follows that it is only the first term in Eq.(\ref{eq:small_q1_form}) that contributes. If  on the other hand  and with  formally $\eps > 1/2$,  it is the second term in Eq.(\ref{eq:small_q1_form}) that contributes in the form of an exponentially growing term which, fortunately,  exactly cancels with a corresponding  exponentially growing contribution  from Eq.(\ref{eq:small_qnorm_form}).  By inspection it is seen that these exclusive limits correspond  to the states $\psi_{l}(\theta)$ or $\psi_{l+1}(\theta)$, respectively. It is only for the exact rational number  $\eps =1/2$ that these two states  lead to the superposition  in Eq.(\ref{eq:lambda_zero}).

 In order to carry out the analysis  in the small $\lambda$-limit in a more careful manner,  we therefore first restrict ourselves  to the values of  $\eps$ in the range $0\leq \eps < 1/2$. We then observe  that the normalization of the state $\psi(\theta)$ using Eq.(\ref{eq:alternative_form}) is always exactly equal to one  when  one keeps  the same order in the parameter  $q$ in both of the  $\vartheta_3[z,q]$-functions in  Eq.(\ref{eq:alternative_form}).  In a  straightforward manner  one then finds, keeping terms in the  $\vartheta_3[z,q]$-functions up to order ${\cal O}(q^8)$,
the following   analytical answer for  $(\Delta \theta)^2$  
 in the small $\lambda$-limit 
\begin{gather} 
(\Delta \theta)^2 =\int_{-\pi}^{\pi} d\theta p_1(\theta)\theta^2  =  \frac{\pi^2}{3} - 4(e^{\displaystyle{-(1-2\eps)/2\lambda}}  +  e^{\displaystyle{-(1+2\eps)/2\lambda}})  \nonumber \\ + e^{\displaystyle{-2(1-\eps)/\lambda}}+  e^{\displaystyle{-2(1+\eps)/\lambda}} 
%-4(e^{\displaystyle{-(5-6\eps)/2\lambda}}  + e^{\displaystyle{-(5+6\eps)/2\lambda}})
+{\cal O}(e^{\displaystyle{-1/\lambda}})   \, ,
\label{eq:wigner2026_1_small_lambda}
\end{gather}
which shows an exponential sensitive to the value of $\eps$.  As $\lambda \rightarrow 0$,  the  uncertainty $(\Delta \theta)^2$  is therefore an increasing  function of $\lambda$, which corresponds to decreasing  values of $\Delta L$  according to  Eq.(\ref{eq:wigner2024_4_2_4}).  

In  the case when $1/2 < \eps  < 1$ we now proceed in a more rigorous manner  as follows. By inspection of  Eq.(\ref{eq:wigner2024_3}),  and apart from  an overall phase,  we notice that $\psi(\theta)$ is exactly invariant for all values of $\lambda$  under the replacement  $\eps \rightarrow -\eps$ provided  that one at the same time makes the change $\psi(\theta) \rightarrow \psi(-\theta)$. In the summation procedure  over integers $n$  in Eq.(\ref{eq:wigner2024_3}) we can then, apart again from  an overall phase,  perform the change $n+ \eps \rightarrow n - (1-\eps)$,  where now   $0 < 1-\eps  < 1/2$.  One can then proceed as in the derivation  of   Eq.(\ref{eq:wigner2026_1_small_lambda})  and obtain the same result re-expressed  in terms  of  $\eps$ in the range $0 < \eps  < 1/2$.

It  is now clear from Eq.(\ref{eq:wigner2026_1_small_lambda}) that the case $\eps=1/2$ plays a special role and must be treated separately.  By making use of the small $q$ expansion $\vartheta_3 [-i \eps/\lambda,q^2] = 2(1+q^4) +{\cal O}( q^{12})$ in  Eq.(\ref{eq:alternative_form}) one now finds that
\begin{gather} 
(\Delta \theta)^2 =\int_{-\pi}^{\pi} d\theta p_1(\theta)\theta^2  =  \frac{\pi^2}{3} -2  -3 e^{\displaystyle{-1/\lambda}} + {\cal O}(e^{\displaystyle{-2/\lambda}})   \, ,
\label{eq:wigner2026_2_small_lambda}
\end{gather}
 in accordance  with the critical value $(\Delta \theta)^2 = \pi^2/3 -2$ in the limit $\lambda \rightarrow 0$ as illustrated  in Fig.\ref{fig:wignerplot_5}.

On the other hand, the asymptotic behaviour of $\Delta\theta$ with $\bar{\theta}=0$  for large values of  $\lambda$,  as exhibited   in Fig.\ref{fig:wignerplot_5}  for the specific case  $\bar{l}= l +1/2$ with $l$ an integer,   can be obtained  by making use of the known asymptotic properties of $\vartheta_3[z,q]$ functions.  For general values of $\eps$  we therefore recall the  following alternative expression for the  state $\psi(\theta)$ according to  Eq.(\ref{eq:wigner2004_5_1}), i.e.,  
\begin{gather} 
\psi(\theta)   
= N e^{ \displaystyle{i(l+ \eps)\theta}}\sum_{n=-\infty}^{\infty} e^{\displaystyle{i2n\pi\eps}}e^{ -\displaystyle{\lambda(\theta + 2n\pi)^2/2}} \, \,  .
%
% \label{eq:wigner2024_3}
\label{eq:alternative_form_2}
\end{gather}
In  Eq.(\ref{eq:second_ alternative_form}) we then consider the expansion
%
%
 %%In Fig.\ref{fig:wignerplot_5} we consider in particular $\eps=1/2$ for  large values of $\lambda$  with $%%%\theta$ in the range  $[-\pi,\pi]$. 
 %%For general values of $\eps$ we  now  make use of the approximation
%
%
\begin{gather} 
\psi(\theta)   
= N e^{ \displaystyle{i(l+ \eps)\theta}} e^{ -\displaystyle{\lambda\theta^2/2}}\Huge( 1 +e^{ -\displaystyle{2\lambda \pi^2}}(e^{ \displaystyle{i2\pi\eps}} e^{ -\displaystyle{2\pi\lambda \theta}}  + e^{ \displaystyle{-i2\pi\eps}}e^{\displaystyle{2\pi\lambda \theta}}) + {\cal O}(e^{\displaystyle{-4\pi^2\lambda}}  ) \Huge)\, \,  ,
\label{eq:alternative_form_3}
\end{gather}
%
%
%in the expression for  the probability distribution $p_1(\theta)=|\psi(\theta)|^2$. 
%%apart from exponentially small corrections of the  order ${\cal O}(\exp(-\lambda\pi^2))$. 
where we make use of 
\begin{gather} 
1= N^2\sqrt{\frac{\pi}{\lambda}} (1 +2e^{\displaystyle{-\pi^2\lambda}}\cos{(2\pi\eps)} + 2e^{\displaystyle{-4\pi^2\lambda}}\cos{(4\pi\eps)}  + {\cal O}(\cos(6\pi\eps)e^{\displaystyle{-9\pi^2\lambda}}) \,\, ,
%= \frac{N^2}{\sqrt{\lambda}} \int_{-\pi\sqrt{\lambda}}^{\pi\sqrt{\lambda}}dx e^{ \displaystyle{-x^2}}
%= (1 + {\cal O}(\frac{e^{\displaystyle{-4\pi^2\lambda}} }{\sqrt{\lambda}})) \, \,  ,
%
\label{eq:alternative_form_5}
\end{gather}
according to  Eq.(\ref{eq:wigner2024_4}).  Therefore  we, e.g.,  find  that $N^2 =\sqrt{\lambda/\pi}$ for sufficiently  large values of $\lambda$ similar to the analysis in  terms of an error function in accordance  with the results in \Referes{Skagerstam_2023,Padgett_2004}.

\begin{comment}
By making use of Eq.(\ref{eq:alternative_form_3}) one then finds for normalization integral for large values of $\lambda$  that
%
%
\begin{gather} 
\int_{-\pi}^{\pi}d\theta p_1(\theta) 
%= \frac{N^2}{\sqrt{\lambda}} \int_{-\pi\sqrt{\lambda}}^{\pi\sqrt{\lambda}}dx e^{ \displaystyle{-x^2}}
= (1  - (\cos{2\pi\eps} + 2\cos{4\pi\eps})\frac{e^{\displaystyle{-5\pi^2\lambda}} }{\sqrt{\lambda\pi^3}}  + % \nonumber \\
 {\cal O}(\frac{e^{\displaystyle{-5\pi^2\lambda}} }{\sqrt{\lambda^3}}) \, \,  ,
 %
 % {\cal O}((\cos{2\pi\eps} + 8\cos{4\pi\eps})\frac{e^{\displaystyle{-5\pi^2\lambda}} }{\sqrt{\lambda^3}}) \, \,  ,
%
%
\label{eq:alternative_form_4}
%
\end{gather}
%
%
For exceptional values of $\eps$ such that $(\cos{2\pi\eps} + 8\cos{4\pi\eps}=0$ the last term in Eq.(\ref{eq:alternative_form_4}) will be smaller with an additional   factor $1/\lambda$.
%
\end{comment}
%
%
%
A more general, systematic and  analytical  method in the  evaluation of $(\Delta \theta)^2$   to any order in  $q$ is based on Eq.(\ref{eq:second_ alternative_form}).  This procedure leads to error functions depending on $\pi\sqrt{\lambda}$  due to the explicit exponential $\theta$-dependence in Eq.(\ref{eq:second_ alternative_form}). For large values of $\lambda$  the asymptotic  expansion of these error functions  then lead to the expression
\begin{gather} 
(\Delta \theta)^2 =\int_{-\pi}^{\pi} d\theta p_1(\theta)\theta^2 
%%= \frac{N^2}{\sqrt{\lambda}} \frac{2}{\lambda} \int_{0}^{\pi\sqrt{\lambda}}dxx^2 e^{ \displaystyle{-x^2}}
 %\rightarrow  \frac{1}{2\lambda} = \frac{1}{4(\Delta L)^2} \, \,  
= \frac{1}{2\lambda}  + 2\pi^2
 e^{\displaystyle{-\pi^2\lambda}}\cos{(2\pi\eps)} (1  - \frac{2}{\sqrt{\lambda \pi^3}})   -   {\cal O}(\frac{e^{\displaystyle{-\pi^2\lambda}}}{\sqrt{\lambda^3}}) %\nonumber \\  + \, \, {\cal O}(\cos{(4\pi\eps)}\frac{e^{\displaystyle{-5\pi^2\lambda}} }{\sqrt{\lambda^3}}) 
 %\frac{e^{\displaystyle{-4\pi^2\lambda}} }{\lambda^3})\,  \, \, 
\,  \, \, ,
 \label{eq:wigner2024_large_lambda}
\end{gather}
%
%
%where $q=\exp(-\lambda\pi)$ 
which shows that the dependence of $\eps$  for large values of $\lambda$ is exponentially suppressed. It turns out that  this expression for $(\Delta \theta)^2$ agrees well with the numerical results as presented in Fig.\ref{fig:wignerplot_5}  for $\eps =1/2$ even for  $\lambda  \geq {\cal O}(0.1)$.
%
%
%%%%%If $\eps$ in Eq.(\ref{eq:wigner2024_large_lambda}) is such that $\cos{2\pi\eps}=0$  then the ${\cal O}( %%%%%%%%e^{\displaystyle{-2\pi^2\lambda}}) $-terms will be absent and exponentially smaller suppressed terms will then %%%contribute.

For large values of $\lambda$ the  approximation  as in Eq.(\ref{eq:alternative_form_3}) can, furthermore be used to find $(\Delta L)^2$ in a tedious but straightforward  manner  with a  result that is consistent with  Eq.(\ref{eq:wigner2024_4_2_3_1}) which was obtained in a different manner.
 By making use  of $(\Delta L)^2 =\lambda/2$  as in Eq.(\ref{eq:wigner2024_4_2_3_1}),  it therefore follows in   large  $\lambda$-limit that  $\Delta \theta \Delta L =1/2$  in accordance  with the results in \Referes{Skagerstam_2023,Padgett_2004}.

 \begin{center}
\subsection{$p_2(\theta)$  and Expansions  of the Parameter $\lambda$ 
\label{subsection_measurements_2}}
\end{center}

If we instead consider  the probability distribution $p_2(\theta)$  for small values of $\lambda$,  we can proceed as in the derivation for Eq.(\ref{eq:wigner2026_1_small_lambda}) and we then obtain the result 
\begin{gather}  
(\Delta \theta)^2 =\int_{-\pi}^{\pi} d\theta p_2(\theta) \theta^2 = \frac{\pi^2}{3} 
% + e^{ -\displaystyle{2(1-\eps)/\lambda}}  + e^{ -\displaystyle{2(1+\eps)/\lambda}} +
 + {\cal O}(e^{ -\displaystyle{1/\lambda}}) \, \,  ,
  \label{eq:wigner2024_small_lambda_02}
 \end{gather}
where we  inherit the condition $0\leq \eps < 1/2$ in order that the distribution $p_1(\theta)$ is well-defined.
In the limiting  case when  $\eps=1/2$ we find in similar manner that
\begin{gather}  
(\Delta \theta)^2 =\int_{-\pi}^{\pi} d\theta p_2(\theta) \theta^2 = \frac{\pi^2}{3}  + e^{ -\displaystyle{1/\lambda}}  - 
 {\cal O}(e^{ -\displaystyle{3/\lambda}}) \, \,  ,
 \label{eq:wigner2024_small_lambda_2}
\end{gather}
which is a decreasing function as $\lambda \rightarrow 0$  as exhibited in Fig.\ref{fig:wignerplot_5}. 

 For large values of $\lambda$  we make use of  the method in the derivation of  Eq.(\ref{eq:wigner2024_large_lambda})  with the result 
%
%
%%\begin{gather} 
%%|\psi(\theta + \pi)|^2=  N^2(e^{ -\displaystyle{\lambda(\theta + \pi)^2}} + e^{ -\displaystyle{\lambda(\theta - \pi)^2}} )  %%+ {\cal O}(e^{ -\displaystyle{\lambda \pi^2}})\, \,  .
%
% \label{eq:wigner2024_3}
%
%%\end{gather}
%
%
%
%
%
\begin{gather} 
(\Delta \theta)^2 =\int_{-\pi}^{\pi}d\theta  p_2(\theta)\theta^2
= \frac{\pi^2}{2} + \frac{1}{2\lambda}  - \sqrt{\frac{\pi}{\lambda}} -   {\cal O}(\frac{e^{\displaystyle{-\pi^2\lambda}}}{\sqrt{\lambda^3}}) \nonumber \\ =  \frac{\pi^2}{2}  +\frac{1}{4(\Delta L)^2} - \frac{1}{\Delta L} \sqrt{\frac{\pi}{2}} -    {\cal O}(\frac{e^{\displaystyle{-\pi^2\lambda}}}{\sqrt{\lambda^3}}) \, \,  ,
 \label{eq:wigner2024_large_lambda_2}
\end{gather}
 where the neglected  terms  are  inherited  from Eq.(\ref{eq:wigner2024_large_lambda}). As in Eq.(\ref{eq:wigner2024_large_lambda})   this expression for $(\Delta \theta)^2$ agrees well with the numerical results as presented in Fig.\ref{fig:wignerplot_5}  for $\eps =1/2$ even for  $\lambda  \geq {\cal O}(0.1)$. For sufficiently large values of $\lambda$, in which case $(\Delta L)^2=\lambda/2$,  it then follows that  $\Delta \theta \Delta L \gg 1/2$.
 % For the exceptional values  $\eps =1/4$ and   $\eps =3/4$ as in Eq.(\ref{eq:wigner2024_large_lambda}),  the terms we %neglect of the order ${\cal O}(q^2)$ with $q=\exp(-\lambda\pi)$ vanishes and will be replaced by exponentially smaller %terms   of order ${\cal O}(q^4)$.

 \begin{center}
\subsection{Characteristic Gap in $\Delta\theta$  for half-integer  values of $\langle L\rangle$ 
\label{subsection_measurements_3}}
\end{center}

We have found a  characteristic  gap $\pi/\sqrt{3} -\sqrt{\pi^2/3 -2}$ for the  observable $\Delta \theta$ for the state in \Refer{Skagerstam_2023} with $\bar{l}= l+1/2$ for any integer $l$ as $\langle L\rangle  \rightarrow l+1/2$ and $\Delta L \rightarrow 1/2$ in the limit  as $\lambda \rightarrow 0$.
Furthermore, we  remark that the approximation  Eq.(\ref{eq:wigner2024_large_lambda})   describes the lower solid curve in Fig.\ref{fig:wignerplot_5} for a wide range of $\lambda$ and that it is  only for $\lambda$ close to zero that Eq.(\ref{eq:wigner2026_1_small_lambda}) is required.  In a similar manner the approximation Eq.(\ref{eq:wigner2024_large_lambda_2})    describes well the upper solid curve in Fig.\ref{fig:wignerplot_5} and that it is only for $\lambda$ close to zero that Eq.(\ref{eq:wigner2024_small_lambda_2}) is required.

For general values of  $\eps \neq 1/2$ in Eq.(\ref{eq:alternative_form}) it now follows from Eqs.(\ref{eq:wigner2026_1_small_lambda}) and  (\ref{eq:wigner2024_small_lambda_02}) that the characteristic gap in Fig.\ref{fig:wignerplot_5} disappears in the limit of small $\lambda$. In this limit and depending on the conditions $0\leq \eps <1/2 $ or  $1/2< \eps <1$,   the upper and lower curves in  Fig.\ref{fig:wignerplot_5} coalesce to the common  point with $\Delta \theta =\pi/\sqrt{3}$ and $\Delta L = 0$ corresponding to dominating pure state contributions $\psi_l(\theta)$ or $\psi_{l+1}(\theta)$, respectively. 
%It can actually be shown analytically as indicated above  that large $\lambda$-limit is only weakly dependent of the %parameter $\eps$ which we also have verified in numerical simulations. 

\begin{center}
\section{CONCLUSIONS}
\label{section_our_conclusions}
\end{center}

In  summary we have considered  two different but related quasi-probability  distributions  $W[\theta,p]$  and    $W_{1/2}[\theta,p]$ as defined in Eqs.(\ref{eq:ourwigner_1}) and (\ref{eq:wigner_kastrup}), respectively, with $\theta$ in the range $-\pi \leq \theta \leq \pi $ and $p$ in principle any real number.        When the parameter  $p$ takes  integer values the corresponding positive marginal distributions  $W[p]$ and  $W_{1/2}[p]$ both agree with expected quantum mechanical probability distributions. The  positive marginal distributions $W[\theta]$ and  $W_{1/2}[\theta]$ obtained by summing over all integers $p$ are properly normalized but different and give rise to different values for the uncertainty $(\Delta \theta)^2$.

 If the range of  $p$ is all real numbers the  distributions $W[\theta,p]$  and    $W_{1/2}[\theta,p]$  differ by a factor $1/2$ in the overall normalization.  We track this difference back to the fact that the different marginal distributions $W[p]$  and    $W_{1/2}[p]$ are in general not positive functions, i.e., they cannot be regarded as probabilities which then make their physical interpretation unclear.  In this sense  the properties in Eq.(\ref{eq:wigner2024_2}) for the Wigner function in Eq.(\ref{eq:wigner2024_1}) can not in general be fulfilled for the $W[\theta,p]$  and    $W_{1/2}[\theta,p]$ for all real numbers $p$. Furthermore, both   distributions show  signs of negativity for fractional values of $p$  in regions which, however, do not necessarily  overlap. If both of them could be measured  for the same quantum state $\psi(\theta)$ the appearance of negativity as a sign of quantum-mechanical  behaviour would then be  ambiguous as illustrated in Figs.\ref{fig:wignerplot_3}  and \ref{fig:wignerplot_4}. 
 
 %Both  distribution $W[\theta,p]$  and    $W_{1/2}[\theta,p]$ lead to  marginal distributions  $W[p]$  and    $W_{1/2}%[p]$ that can be negative in different regions   which then make their physical interpretation unclear. 
 
 % It is of importance to observe  one of several  special features of the distribution  $W_{1/2}[\theta,p]$  is such  that %if it  could be determined experimentally for integers $p$  that would, nevertheless, be sufficient in order to infer the %marginal distribution $W_{1/2}[\theta]=|\psi(\theta)|^2$. One could then  find   $|\psi(\theta  \pm \pi)|^2$  by  a %periodic extension as well as the marginal distribution $W[\theta]  = ( |\psi(\theta)|^2 +  |\psi(\theta + \pi)|^2 )/2$.
  
   From a knowledge of the explicit form of the  probabilities  $p_1(\theta)$ and $p_2(\theta)$ as given  in Eqs.(\ref{eq:half_probability}) and (\ref{eq:whole_probability}), respectiveley,  one  obtains  the corresponding uncertainties $\Delta \theta$. For the observable  $\Delta L$  in the case  $\bar{l} = l+ 1/2$  with integers $l$,  we therefore predict the behaviour as presented in Fig.\ref{fig:wignerplot_5} using states of the form  $\psi(\theta)$ in \Refer{Skagerstam_2023}.  Here the uncertainties $\Delta \theta$ and $\Delta L$ are, in particular,  independent of the integer  $l$ if $\bar{\theta} = 0$. A specific gap in $\Delta \theta$ is predicted depending  on which marginal distribution $p_1(\theta)$  or $p_2(\theta)$  is used in order  to obtain the uncertainty $\Delta \theta$.

\begin{comment}
Due to the explicit definitions of the quasi-probability  distributions in Eqs.(\ref{eq:ourwigner_1})  and (\ref{eq:wigner_kastrup})     the corresponding integration procedures  determines the  effective range of  the  $\theta$-dependence of the states $\psi(\theta)$ considered.   The   distribution $W[\theta,p]$  therefore extends this   effective  range  for the states $\psi(\theta)$ to $[-2\pi,2\pi]$  in comparison with  the  corresponding effective $\theta$-dependence $[-3\pi/,3\pi/2]$ 
for the distribution $W_{1/2}[\theta,p]$.  Expressed in terms of the state-dependent  probability distributions 
 $p_{1/2} (\theta)$ and $p(\theta)$  in Eqs.(\ref{eq:half_probability}) and (\ref{eq:whole_probability}), respectively, $p_{1/2} (\theta)$ has an effective  range $\theta\in[-\pi,\pi]$  and  $p(\theta)$ an  effective range $\theta\in[-2\pi,2\pi]$  for the    $\theta$-dependence of the same state $\psi(\theta)$. 
\end{comment}

For a half-integer angular momentum eigenstate    well-known rules of quantum mechanics imply that a $4\pi$-rotation  is usually required in order to obtain the same eigenstate and  leads to  unique physical consequences. In our case with half-integer expectation values of $\langle L \rangle$ and  for  fractional values of $p$ it is therefore not obvious why, e.g.,  the  quasi-probability  distribution   $W[\theta,p]$, with a range $[-2\pi,2\pi]$ of integration in Eq.(\ref{eq:ourwigner_1}),  can,  as noticed in the present work,  exhibit a more pronounced  negativity  than quasi-probability  distribution $W_{1/2}[\theta,p]$, with a range $[-\pi, \pi]$ of integration in Eq.(\ref{eq:wigner_kastrup}).  In both cases the half-integer value   $p=1/2$ appears, however,  to be  of special importance.
 
 To what extent  quasi-probability  distributions, like $W[\theta,p]$  or $W_{1/2}[\theta,p]$ in Sections \ref{section_ourW}  and \ref{section_polish_W}, respectively, turn out to be useful  in the analysis of pure quantum states in, e.g.,  the form as  considered in the present work in Section \ref{section_our_state} can, of course, only be determined in terms  of  actual   experimental situations. This has been emphasized  in, e.g.,  \Refer{Agarwal_1994} concerning, in particular,  the physics that involves the notion of a quantum phase, a concept which also plays an important role in the present work. 
 
With regard to  Section \ref{section_our_uncertainties} in the present paper,   discussions on fundamental quantum-mechanical limitations of measurement of fluctuations like $\Delta \theta$ and $\Delta L$ can be found in, e.g.,  Refs.\cite{Clerk_2010,DiMario_2020,Knoll_2025} and references cited therein.
%
% ---------------------   acknowledement  (start) --------------------------
%\newpage
\vspace{0.5cm}
\begin{center}
   {\bf \large ACKNOWLEDGEMENT}
\end{center}
\vspace{0.5cm}
\noindent The research  for B.-S. Skagerstam (B.-S.S)   has been supported by NTNU at Trondheim  and Molde University College and for  P.K. Rekdal   by  Molde University College. We dedicate this work to the memory of John R. Klauder (1932-2024) who played an important role in discussions with B.-S.S. on problems related to  topics of the present work.  B.-S.S. is  also grateful to in particular S. Shabanov   and the University of Florida at Gainesville for the hospitality  in connection with  the   wonderful  {\sl  John R. Klauder Memorial Conference} held at the Department of Mathematics, February 15, 2025, when  the present work was  in progress. Finally, the authors are grateful for various constructive comments and suggestions  from  colleagues and anonymous  readers on the first version of this work.
%
% ---------------------   acknowledement  (end) -------------------------
%
%
%
%\end{comment}
%
%
%
%
% ---------------------   bibliography  (start) --------------------------
% 
%\newpage
%
\vspace{0.5cm}
\begin{center}
   {\bf \large REFERENCES}
\end{center}
\vspace{0.25cm}
%
 
%
%
% ---------------------   bibliography  (end) --------------------------
% 
%
\end{document}